\documentclass[10pt,final, twocolumn]{IEEEtran}
\IEEEoverridecommandlockouts
\usepackage[caption=false,font=normalsize,labelfont=sf,textfont=sf]{subfig}
\usepackage{cite}
\usepackage{amsmath,amssymb,amsfonts,bm}
\usepackage{graphicx}
\usepackage{textcomp}
\usepackage{xcolor}
\usepackage{amsmath,amssymb,amsfonts,graphicx,mathtools,nccmath,amsthm,float}
\usepackage{multicol,tikz,cite,array,booktabs,url}
\usepackage{cite}
\usepackage{textcomp}
\usepackage{xcolor}
\usepackage{algorithm}
\usepackage{algpseudocode,algorithmicx}
\usepackage{comment}
\usepackage[top=.75in, left= 0.625in, right = 0.63in, bottom =1in]{geometry}
\def\BibTeX{{\rm B\kern-.05em{\sc i\kern-.025em b}\kern-.08em
    T\kern-.1667em\lower.7ex\hbox{E}\kern-.125emX}}
    
\algnewcommand{\LineComment}[1]{\State \(\#\) #1}
    
\graphicspath{{figs/}}
\DeclareGraphicsExtensions{.pdf}

\newtheorem{Lemma}{Lemma}
\newtheorem{Prop}{Proposition}

\DeclarePairedDelimiterX\set[1]\lbrace\rbrace{#1}

\DeclarePairedDelimiter\floor{\lfloor}{\rfloor}

\begin{document}

\title{Near-Field Beam Tracking with\\ Extremely Large Dynamic Metasurface Antennas}
\author{Panagiotis~Gavriilidis,~\IEEEmembership{Graduate Student Member,~IEEE} and George C. Alexandropoulos,~\IEEEmembership{Senior Member,~IEEE} 
\thanks{This work has been supported by the SNS JU project TERRAMETA under the European Union’s Horizon Europe research and innovation programme under grant agreement number 101097101, including top-up funding by UKRI under the UK government's Horizon Europe funding guarantee.}
\thanks{The authors are with the Department of Informatics and Telecommunications,
National and Kapodistrian University of Athens, 15784 Athens, Greece (e-mails: \{pangavr, alexandg\}@di.uoa.gr).}
}



\maketitle
\begin{abstract}
The interplay between large antenna apertures and high frequencies in future generations of wireless networks will give rise to near-field communications. In this paper, we focus on the hybrid analog and digital beamforming architecture of dynamic metasurface antennas, which constitutes a recent prominent enabler of extremely massive antenna architectures, and devise a near-field beam tracking framework that initiates near-field beam sweeping only when the base station estimates that its provided beamforming gain drops below a threshold from its theoretically optimum value. Novel analytical expressions for the correlation function between any two beam focusing vectors, the beamforming gain with respect to user coordinate mismatch, the direction of the user movement yielding the fastest beamforming gain deterioration, and the minimum user displacement for a certain performance loss are presented. We also design a non-uniform coordinate grid for effectively sampling the user area of interest at each position estimation slot. Our extensive simulation results validate our theoretical analysis and showcase the superiority of the proposed near-field beam tracking over benchmarks.
\end{abstract}


\section{Introduction} \label{Sec:Intro}
\IEEEPARstart{H}{igh} frequency wireless systems are expected to play a prominent role in the upcoming sixth Generation (6G) of mobile networks, offering extensive bandwidth that can enable ultra-high data rate communications and high precision localization and sensing~\cite{THz_loc_tutorial}. To compensate large signal propagation losses at millimeter Wave (mmWave) and beyond, massive, and very recently, holographic Multiple-Input-Multiple Output (MIMO) systems are being investigated~\cite{Holographic_MIMO_et_al}. The latter transceivers incorporate extremely large numbers of densely packed antenna elements capable of realizing highly directive beams towards possibly massive numbers of mobile users~\cite{XLMIMO_tutorial}.

The combination of high frequencies and extremely large MIMO brings forth near-field wireless communications~\cite{NF_tutorial}, where the curvature of the spherical wavefront is non-negligible. Such systems are expected to drastically increase connectivity, by extending angle division multiple access \cite{Hybrid_mimo_tracking} to location division multiple access \cite{SDMA_vs_LDMA}, and enable high accuracy localization and sensing even with narrowband signals, as compared to schemes relying on time of arrival estimation requiring wideband signaling~\cite{time_difference_arrival_localisation,RIS_aided_time_of_arrival,Localization_TOA_Primer}. However, conventional fully digital MIMO and even hybrid analog and digital architectures, contain power-hungry Radio Frequency (RF) chains and networks of phase shifters that prohibit their scaling to extremely large numbers of antennas. To deal with this issue, the concept of Dynamic Metasurface Antennas (DMAs) has been lately introduced~\cite{DMA_Magazine}, which envisions transceivers with extremely large planar arrays of densely packed metamaterials with programmable responses. DMA transceivers constitute a special case of Reconfigurable Intelligent Surfaces (RISs)~\cite{RIS_overview} that are equipped with transmit/receive RF chains attached via waveguides to the metamaterial panels, realizing power-efficient hybrid analog and digital beamforming. Because of their attractive features, including the ease of massively scaling the number of metamaterials resulting in advantageous implications on RF chain design~\cite{stacked_hmimo,DMA_1bit_uplink}, DMAs are lately receiving increased research and development attention~\cite{gavras2023duplex,DMA_UL_mMIMO,DMA_near_field_channel,DMA_loc_Nir,DMA_energy_eff,HMIMO_survey_et_al}.

In high frequency massive MIMO, and beyond, wireless systems, the radiative near-field region is extended. This enables profiting from the curvature of the signal's wavefront to extend direction-of-arrival estimation to localization~\cite{Near_Field_localization_MUSIC}. Among the first works that analyzed the Cram\'{e}r-Rao Bound (CRB) for source localization in the near field belongs to~\cite{CRB_source_loc}. In~\cite{NF_tracking}, the authors derived the posterior CRB and the Fisher Information Matrix (FIM) for near-field localization, taking into account a priori knowledge, thus, tailoring their analysis to a tracking scenario. User tracking in an RIS-aided wireless setup was investigated in~\cite{RIS_and_NF_tracking}, where an extended Kalman filtering approach was presented together with an optimization method for the joint design of the precoding matrix at the Base Station (BS) and the RIS phase configuration that improves localization. A framework for maximizing the likelihood estimate of the channel gain and the User Equipment (UE) coordinates, through a FIM-based approach, was designed in~\cite{RIS_localisation_George_henk_ML_estimator}. Very recently, the authors in~\cite{DMA_loc_Nir} considered a DMA-based BS and presented a maximum likelihood localization framework. 

The implementation of high directivity in the near-field enables orthogonality even when signals point towards the same angular direction. However, there exist limits imposed on the orthogonality of the beam focusing vectors, which are based on the distance of the UE/target from the antenna array and on the antenna's physical dimensions~\cite{Ramezani2024}. The authors in~\cite{Bjornson_dist} were among the first to study the depth of focus limits for planar arrays, considering a receiver along the normal vector originating from the center of the array. The correlation function of a focusing vector was derived in~\cite{Appendix_approx} for uniform linear arrays, and later, this analysis was extended to the case of uniform planar arrays~\cite{SDMA_vs_LDMA}. In the latter work, a spherical-domain codebook for beamforming was also designed. Very recently, in~\cite{Near_field_NOMA}, the impact of limited resolution in beam focusing vectors was analyzed and, capitalizing on this analysis, a framework for non-orthogonal multiple access was presented.

In this paper, we study the near-field beam tracking problem between a BS equipped with a DMA transceiver and a mobile single-antenna UE in high frequency wireless communication channels. Differently from the previous art and targeting realistic BS deployments, we focus on a generic scenario where the mobile UE, to be dynamically beam focused, lies in a plane vertical to the BS plane with a constant height difference. The paper's contributions are summarized as follows:
\begin{itemize}
    \item The correlation function between any two beam focusing vectors in the near-field region with the considered DMA-based transceiver architecture and deployment scenario, is analyzed. A novel analytical approximation for this function, which is decoupled in terms of the parameters of the range and angle of the focusing point, is presented.  
    \item Novel analytical results extending the near-field effects to larger distances, even larger than the Rayleigh distance, and expressions for the depth of focus limits are derived.
    \item We present novel analytical expressions for the direction of the UE movement that yields the fastest beamforming gain deterioration, as well as for the minimum UE displacement needed so that a certain percentage of the optimum beamforming gain is lost due to beam misfocusing.
    \item A novel non-uniform coordinate grid for effectively sampling the UE area of interest is designed. This grid is dynamically reconfigured at each position estimation slot.
    \item We present a near-field beam tracking framework according to which the BS performs beam focusing sweeping when it estimates that its provided beamforming gain drops below a threshold from its optimum value. We introduce the metric of the effective beam coherence time, indicating the minimum time needed for the UE to experience a specific loss relative to the theoretically optimal beam focusing gain, that is dynamically computed at the BS during each UE position estimation slot. 
\end{itemize}
\textit{Notations:} Vectors and matrices are represented by boldface lowercase and boldface capital letters, respectively. $\mathbf{I}_{n}$ and $\mathbf{0}_{n\times 1}$ ($n\geq2$) are the $n\times n$ identity matrix and the \(n \times 1\) zeros' vector, respectively. $[\mathbf{A}]_{i,j}$ is the $(i,j)$-th element of $\mathbf{A}$, $[\mathbf{a}]_i$ and $||\mathbf{a}||_2$ denote $\mathbf{a}$'s $i$-th element and Euclidean norm, respectively, and \(|\cdot|\) gives the amplitude (absolute value) of a complex (real) scalar. $E[\cdot]$ is the expectation operator and $\mathbf{x}\sim\mathcal{CN}(\mathbf{a},\mathbf{A})$ indicates a complex Gaussian random vector with mean $\mathbf{a}$ and covariance matrix $\mathbf{A}$, and $\jmath\triangleq\sqrt{-1}$ is the imaginary unit. Finally, \({\rm acos}(\cdot)\) is the arc cosine function. 

\section{System and Channel Models} \label{Sec:Sys_Model}
\begin{figure}
    \centering
    \includegraphics[scale = 0.75]{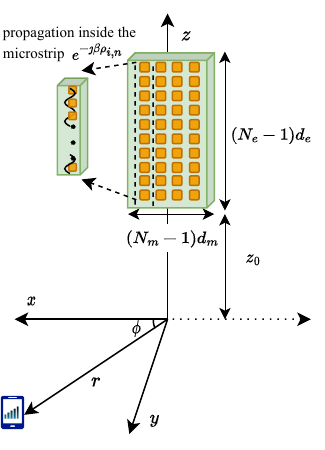}
    \caption{The considered system model comprising a static BS equipped with a DMA of an extremely large number of metamaterials, lying in the \(yz\) plane and a mobile single-antenna UE moving inside the \(xy\)-plane.}\vspace{-3mm}
    \label{fig:System_model}
\end{figure}

Consider a high frequency point-to-point wireless communication system between a multi-antenna BS and a mobile single-antenna UE (see Fig.~\ref{fig:System_model}), where the former node is equipped with a DMA-based transceiver consisting of \(N_m\) single-RF-fed microstrips, each including \(N_e\) metamaterial elements; the total number of BS antenna elements is thus \(N \triangleq N_m N_e\). It typically holds that \(N_e\gg N_m\), indicating the advantage of this metasurface-based antenna architecture to easily incorporate extremely massive numbers of low-cost metamaterials per microstrip~\cite{DMA_Magazine}. Due to the considered high frequencies (i.e., mmWave and beyond) and the large DMA aperture at the BS, we focus on the near-field region where mainly the BS-UE communication is expected to take place. The Rayleigh distance \(2D^2\lambda^{-1}\), where \(D\) denotes the array aperture and \(\lambda\) the signal propagation wavelength, defines the limit between the near- and far-field regions of wave propagation~\cite{NF_tutorial}.

In the coordinate system of Fig.~\ref{fig:System_model}, \(\phi \in [0,\pi]\) denotes the azimuth angle and \(r\) is the distance of the UE from the origin, with this node assumed to move inside a plane vertical to the plane of the BS's DMA. The BS is assumed to be placed at a height represented by \(z_0\) in the \(z\)-axis from the UE plane. In the following Section~\ref{sec:BF_analysis}, capitalizing on this coordinate system, we will present a novel near-field beamforming method and analysis, which would be otherwise infeasible due to the existence in the range expression formula \cite[eq.~(10)]{SDMA_vs_LDMA} of a bilinear term, as well as the beamdepth's dependency to both the azimuth and elevation angles~\cite[eq.~(26)]{SDMA_vs_LDMA}.
Let \(d_e\) and \(d_m\) denote respectively the inter-element and inter-microstrip distances in Fig.~\ref{fig:System_model}. Then, the Euclidean distance between each \(n\)-th metamaterial (\(n=0,1,\ldots,N_e-1\)) of each \(i\)-th microstrip (\(i=0,1,\ldots,N_m-1\)) at the BS and the UE located at the point\footnote{We henceforth make use of the compact notation \(\mathbf{p}=[r,\phi]\) for the UE position for a given height difference \(z_0\) from the BS.} \(\mathbf{p}=[r\cos(\phi),r\sin(\phi),0]\) can be expressed as follows:
\begin{align}\label{eq:Dist_approx}
    &r_{i,n}\!=\!\bigg( \!\!\left(r\cos(\phi)\!-\!i_xd_m \right)^2\!+\!\left(r\sin(\phi)\right)^2\!+\! \left(nd_e\!+\!z_0\right)^2\!\!\bigg)^{\frac{1}{2}}   \\
   & \approx\!r +\!\frac{i_x^2d_m^2}{2r}(1\!-\!\cos^{2}(\phi))\!-\!\cos(\phi)i_x d_m\!+\!\frac{n^2 d_e^2}{2 r}\!+\!\frac{z_0 n d_e}{r}\!+\!\frac{z_0}{2r},\nonumber
\end{align}
where the \((i,n)\)-th DMA element is located at the point \([i_{x} d_m, 0, nd_e+z_0]\) with \(i_{x}\triangleq i - 0.5(N_m-1)\). For the second's row approximation, we have used the Taylor expansion \(\sqrt{1+x}\approx 1 +\frac{x}{2}-\frac{x^2}{8}\) neglecting the terms divided by \(r^2\), as per the \textit{Fresnel approximation} \cite{NF_tutorial}. However, in our coordinate system, the BS does not serve as the origin, hence, for the maximum phase difference to be less than \(\frac{\pi}{8}\) at \(\phi = \frac{\pi}{2}\), we employ the Lagrange error bound \cite{apostol1991calculus}, yielding the expression: 
\begin{align}\label{eq:r_bound}
    r\geq r_{\rm appr} \triangleq \left(\frac{2L^4_{z_0}}{\lambda}\right)^{1/3},
\end{align}
where \(L_{z_0}\triangleq\sqrt{((N_e-1)d_e + z_0)^2 + 0.25(N_m-1)^2 d^2_m}\).
Equivalently, for \(r_0\) being the distance from the DMA's center, i.e., \(r_0 = \sqrt{r^2 +(z_0+0.5(N_e-1))^2 }\), the approximation in \eqref{eq:Dist_approx} holds for \(r_0 \geq r_{0,{\rm appr}} \triangleq \sqrt{r_{\rm appr}^2 +(z_0+0.5(N_e-1))^2} \). It is noted that this result aligns with the literature when bringing the coordinate system to the DMA's center; this is accomplished by setting \(z_0= -0.5(N_e-1)d_e\). For this origin shifting, the approximation holds for \(r_0 \geq \frac{D}{2} \left(\frac{D}{\lambda}\right)^{1/3}\)~\cite[eq.~(19)]{Rayleigh_Fresnel_distances}. 
\subsection{DMA Modeling}
To compensate for the high propagation loss at mmWave, and beyond, frequencies, we assume that the BS adopts hybrid analog and digital beamforming~\cite{HMIMO_survey_et_al}, which is a channel-dependent technique necessitating the availability of Channel State Information (CSI). However, the BS-UE channel gain vector is both difficult and time-consuming to acquire due to the extremely massive number of antennas at the BS, a fact that has motivated spatially sparse channel representations~\cite{vlachos2019wideband,Wideband_Hybrid_Tracking}. We particularly focus, in this paper, on the DownLink (DL) direction targeting beam focusing toward the UE position, i.e., to achieve the maximum possible gain at this location. Let $s$ denote the unit-power complex-valued information symbol transmitted via the beamformed vector \(\mathbf{x} \triangleq \mathbf{\bar{Q}}\mathbf{v}s\in \mathbb{C}^{N \times 1}\), where $\mathbf{\bar{Q}}\in \mathbb{C}^{N \times N_m}$ and \(\mathbf{v}\in \mathbb{C}^{N_m \times 1}\) represent the DMA analog weights and the digital beamformer at the BS, respectively. The transmitted vector is assumed power limited as \(||\mathbf{x}||^2_2\leq P_b\) with \(P_b\) being the maximum BS transmit power. The DMA configuration is defined \({\mathbf{\bar{Q}}} \triangleq \mathbf{P}_{m}\mathbf{Q}\), where the diagonal matrix \(\mathbf{P}_m\in \mathbb{C}^{N \times N}\) models signal propagation inside each microstrip and \(\mathbf{Q} \in \mathbb{C}^{N \times N_m}\) includes the tunable responses of the identical metamaterials~\cite{DMA_near_field_channel}. More specifically, for each \(n\)-th element of each \(i\)-th microstrip, and assuming a lossless waveguide at each microstrip, the phase distortion due to the intra-microstrip propagation is modeled as follows:
\begin{equation}\label{eq:P_matrix}
 [\mathbf{P}_m]_{(i-1)N_e+n, (i-1)N_e+n}\triangleq \exp\left(-\jmath \beta\rho_{i,n}\right),
\end{equation}
with \(\beta \triangleq 2 \pi\lambda^{-1}\sqrt{\epsilon}\) being the microstrip’s wavenumber and \(\epsilon\) its dielectric constant, and \(\rho_{i,n}\) denotes the distance of the \(n\)-th element in the \(i\)-th microstrip from the
output port. Finally, the elements of the analog beamformer $\mathbf{Q}$ are assumed to follow a Lorentzian-constrained phase model, according to which:
\begin{equation}\label{eq:Q_matrix}
[\mathbf{Q}]_{(i-1)N_{e}+n,j}=\begin{cases} q_{i,n}\in \mathcal{Q},&i=j\\0,&i\neq j\end{cases}
\end{equation}
with $\mathcal{Q}\triangleq\left\{0.5\left(\jmath + e^{\jmath\phi}\right) | \phi \in \left[0,2\pi\right]\right\}$.

\subsection{Channel Model and Received Signals}
In the near-field regime, the celebrated steering vector of the far-field turns into the focusing vector that also captures the effects of the range parameter. To this end, the Line-of-Sight (LoS) vector \(\mathbf{h}_{\text{LoS}}\in \mathbb{C}^{N \times 1}\) between the BS and UE is modeled as follows~\cite{NF_tutorial}: 
\begin{align}\label{eq:LoS_TX_to_UE_channel}
    [\mathbf{h}_{\text{LoS}}]_{(i-1)N_e+n}\triangleq \frac{\lambda}{4 \pi r_{i,n}} e^{-j \frac{2 \pi}{\lambda} r_{i,n}},
\end{align}
where \(r_{i,n}\) denotes the distance from each \(n\)-th element in each \(i\)-th microstrip of the DMA to the UE. Accounting for potential scatterers in the environment (i.e., Non-LoS (NLoS) components), the overall channel vector is given by:
\begin{equation}\label{eq:TX_to_UE_channel}
    \mathbf{h}\triangleq \mathbf{h}_{\rm LoS}+\sum_{\ell=1}^{L}\mathbf{h}_{{\rm NLoS},\ell},
\end{equation}
where \([\mathbf{h}_{{\rm NLoS},\ell}]_{{(i-1)N_e+n}} \triangleq g_{\ell}\frac{\lambda}{4 \pi r_{\ell,i,n}} e^{-j \frac{2 \pi}{\lambda} r_{\ell,i,n}} \) with \(g_{\ell} \triangleq e^{-\jmath w}e^{-\jmath \frac{2\pi}{\lambda} d_{\ell}} \frac{\lambda }{4 \pi d_{\ell}}\), whereas \(d_{\ell}\) and \(\,r_{\ell,i,n}\) denote the distances from the UE to \(\ell\)-th scatterer and from the \(\ell\)-th scatterer to the BS, respectively, as per the linear reflection model \cite{RIS_energy_eff_et_al},  and \(w \sim \mathcal{U}(-\pi,\pi)\) models the reflection-induced phase shift. In this paper, we focus on the radiative near-field region where \(r_{\rm FD}\leq r_{0} \leq r_{\rm RD}\) with \(r_{\rm FD} \triangleq 0.62 \sqrt{D^3\lambda^{-1}}\) and \(r_{\rm RD} \triangleq2 D^2\lambda^{-1}\) denoting the Fresnel and Rayleigh distances, respectively \cite{Rayleigh_Fresnel_distances}. In this region, the pathloss term over the DMA's aperture can be considered as constant~\cite{NF_tutorial}, thus the LoS channel in \eqref{eq:LoS_TX_to_UE_channel} can be reformulated as \(\mathbf{h}_{\rm LoS}=\frac{\lambda}{4 \pi r_{0}}\mathbf{a}(r,\phi)\) with \([\mathbf{a}(r,\phi)]_{(i-1)N_e+n}\triangleq e^{-j \frac{2 \pi}{\lambda} r_{i,n}} \) being the focusing vector.

The narrowband received signals in baseband representation at the UE, in the DL direction, and at the outputs of the $N_m$ microstrips of the BS, in the UpLink (UL) direction, are given, respectively, as follows:
\begin{align}
     y_u\triangleq  & \mathbf{h}^{\rm H}\mathbf{x} + n_u, \label{eq:Rx_UE}\\
    \mathbf{y}_b\triangleq &\mathbf{\bar{Q}}^{\rm H}\mathbf{h}x_u + \mathbf{\bar{Q}}^{\rm H} \mathbf{n}_b, \label{eq:DMA_UE}             
\end{align}
where \(n_u \sim\mathcal{CN}(0,\sigma^2)\) and \(\mathbf{n}_b\sim\mathcal{CN}(\mathbf{0}_{N\times 1},\sigma^2\mathbf{I}_{N}) \) represent the Additive White Gaussian Noise (AWGN) contributions at the UE and BS, respectively, and \(x_u\triangleq \sqrt{P_u}\), with \(P_u\) being the UE's transmit power, denotes the pilot symbol transmitted from the UE in the UL to enable the node's position estimation at the BS side, as will be discussed in the sequel.

\section{Near-Field Beamforming Analysis}\label{sec:BF_analysis}
In this section, we first describe the optimum DMA-based near-field beamformer for the case of availability of the UE position coordinates. Then, novel analytical expressions for the relative beamforming gain under mismatch on the UE position are presented. Finally, we introduce the effective beam coherence time metric, which will be leveraged in Section~\ref{sec:Proposed_BA} within our proposed near-field beam tracking framework.

\subsection{Beamforming Optimization for a Given UE Position}
The analog beamformer at the DMA that maximizes the signal-to-noise ratio at a specific UE position \(\mathbf{p} = [r,\phi]\) is derived via phase-aligning \(\mathbf{h}_{\rm LoS}^{\rm H}\)~\cite[Lemma 1]{DMA_loc_Nir}, i.e., by setting \(q_{i,n}=0.5\left(\jmath + e^{\jmath(\angle{[\mathbf{a}(r,\phi)}]_{(i-1)N_e+n}+\beta \rho_{i,n})}\right)\) $\forall$$i,n$. When the only available channel knowledge are the UE position coordinates, the latter beamformer corresponds to the optimal one. Let us define the beamforming gain as \( G \triangleq |\mathbf{a}^{\rm H}(r,\phi)\mathbf{\bar{Q}}\mathbf{v} |^{2}\), which corresponds to the gain achieved at the UE's position $\mathbf{p}$.
\begin{Prop}\label{prop1}
By setting the DMA analog beamforming weights as \(q_{i,n}=0.5\left(\jmath + e^{\jmath(\angle{[\mathbf{a}(r,\phi)}]_{(i-1)N_e+n}+\beta \rho_{i,n})}\right)\) $\forall$$i,n$ (consequently, the analog beamformer $\mathbf{Q}$ via \eqref{eq:Q_matrix}) and optimizing \(G\) with respect to the DMA digital beamformer \(\mathbf{v}\), the optimum beamforming gain is obtained as \(G_{\rm opt} = 0.5\, P_b N_m N_e\).
\end{Prop}
\begin{proof}
We first compute $\forall$$i=0,1,\ldots,N_m-1$:
    \begin{align*}
        & [\mathbf{a}^{\rm H}(r,\phi)\mathbf{\bar{Q}}]_{i}=\sum_{n=0}^{N_e-1} q_{i,n} [\mathbf{a}(r,\phi)]^{\ast}_{(i-1)N_e+n} e^{-\jmath \beta \rho_{i,n}} = 0.5 N_{e} +\\
        &   0.5 \jmath  \underbrace{\sum_{n=0}^{N_e-1} e^{-\jmath \frac{2 \pi}{\lambda}\left( n \sqrt{\epsilon} d_e -\frac{z_0 n d_e}{2 r}-\frac{n^2 d_e^2}{2r}   - r - \frac{i_x^2 d_m^2\sin^2(\phi)}{2 r} + i_x d_m \cos(\phi)\right)}}_{\triangleq W_i} 
    \end{align*}
For far-field communications, it holds \(|W_i|=\frac{|\sin(N_e\frac{\pi d_e \sqrt{\epsilon}}{\lambda})|}{|\sin(\frac{\pi d_e \sqrt{\epsilon}}{\lambda})|}\), and for the values of \(d_e>\frac{\lambda }{\sqrt{\epsilon} N_e}\) it holds that \(\frac{|W_i|}{N_e} \to 0\). In the regime of near-field communications, \(|W_i|\) still tends to zero since there is no focus at a specific range \(r\)~\cite{SDMA_vs_LDMA}. The latter statement will be made clearer when dealing with the beamforming gain loss with respect to a mismatch in \(r\) (see Lemma~\ref{prop_Dr}). As a result, \([\mathbf{a}^{\rm H}(r,\phi)\mathbf{\bar{Q}}]_{i}\approx 0.5 N_e \, \forall i\). In continuance, the optimal digital beamformer \(\mathbf{v}\) is chosen to solve the optimization problem:
\begin{align*} 
    \mathcal{OP}_1: \,\, \mathbf{v^{\rm opt}}\triangleq\max_{\mathbf{v}} \,\,|\mathbf{a}^{\rm H}(r,\phi)\mathbf{\bar{Q}}\mathbf{v} |^{2} \quad \text{s.t.} \quad ||\mathbf{\bar{Q}v}||^{2}_{2} \leq P_b.
\end{align*}
By careful inspection of $\mathcal{OP}_1$'s objective, it is evident that it can be written as a Generalized Rayleigh Quotient (GRQ). Let us first define: \(\mathbf{A} \triangleq \left(\mathbf{a}^{\rm H}(r,\phi)\mathbf{\bar{Q}}\right)^{\rm H}\mathbf{a}^{\rm H}(r,\phi)\mathbf{\bar{Q}}\), \(\mathbf{B}\triangleq \mathbf{\bar{Q}^{\rm H}\bar{Q}} \approx 0.5 N_e\mathbf{I}_{N_m}\)  (i.e., \([\mathbf{B}]_{i,i}=0.5 N_e + {\rm Im}[W_i^{\ast}]\approx 0.5 N_e\) $\forall$$i$), and \(\mathbf{f}\triangleq \mathbf{B}^{1/2} \mathbf{v}\). Following the GRQ theorem, \(\mathbf{f}^{\rm opt}\) is given by the principal eigenvector of \(\mathbf{B}^{-1/2}\mathbf{A}\mathbf{B}^{-1/2}\), which is a real-valued matrix of only ones, yielding \(\mathbf{f}^{\rm opt} = \sqrt{P_b/N_m} [1,1,\dots,1]^{\rm T}\). Finally, \(\mathbf{v}^{\rm opt}=\mathbf{B}^{-1/2}\mathbf{f}^{\rm opt}= \sqrt{\frac{2 P_b}{N_m N_e}} [1,1,\dots, 1]^{\rm T}\) which yields the beamforming gain \(G_{\rm opt}= 0.5 P_b N_m N_e\).
\end{proof}

Following Proposition~\ref{prop1}, it can be easily computed that \(\mathbf{\bar{Q}}\mathbf{v}_{\rm opt}= 0.5\sqrt{\frac{2 P_b}{N}}\left(\mathbf{a}(r,\phi) + \jmath \,e^{-\jmath \beta \boldsymbol{\rho}}\right)\), representing the maximum ratio transmission vector for the case of a perfectly known UE position $\mathbf{p}$ (i.e., the DMA optimal hybrid analog and digital precoder). The term \(e^{-\jmath \beta \boldsymbol{\rho}}\) models waveguide propagation within the microstrips and does not contribute to the beamforming gain, as shown in the proposition. Note that, when the precoder is not optimal, i.e., when there is a mismatch between \(\mathbf{p}\) and the focusing position \(\mathbf{\hat{p}}\triangleq[\hat{r},\hat{\phi}]\), the beamforming gain is obtained as \(G_{\hat{r},\hat{\phi}}\triangleq 0.5 \frac{P_b}{N} |\mathbf{a}^{\rm H}(r,\phi)\mathbf{a}(\hat{r},\hat{\phi}) |^2\). Consequently, the relative beamforming gain (or correlation factor), defined as the fraction of  \(G_{\hat{r},\hat{\phi}}\) divided by \(G_{\rm opt}\), is given by  \(\frac{G_{\hat{r},\hat{\phi}}}{G_{\rm opt}}\triangleq \frac{|\mathbf{a}^{\rm H}(r,\phi)\mathbf{a}(\hat{r},\hat{\phi}) |^{2}}{N^2 } \). We will next characterize analytically this relative beamforming gain with respect to the difference between the true UE position and the position actually illuminated via the BS's DMA-based beam focusing.

\subsection{Beamforming Gain under UE Coordinate Mismatches}\label{suubsec:BF_gain_mmismatch}
We begin by investigating the beamforming loss due to a range mismatch, i.e., when \(\hat{r}=r\pm \Delta r\) with \(\Delta r \geq 0\).

\begin{Lemma}\label{prop_Dr}
When the BS beam focuses at the point \((\hat{r}, \phi)\), the relative beamforming gain is obtained as~\(\frac{G_{\hat{r},\phi}}{G_{\rm opt}}\triangleq\frac{|\mathbf{a}^{\rm H}(r,\phi)\mathbf{a}(r \pm \Delta r,\phi) |^{2}}{N^2 } = \mathcal{K}^2\left(a(\pm \Delta r), \phi\right)\), where function \( \mathcal{K}\left(x, \phi\right) \triangleq I(x)\left(1-\frac{\pi^2}{90}\left(x \frac{(N_m-1) d_m}{2(N_e-1) d_e}|\sin(\phi)|\right)^4\right)\) 
with \(a(x) \triangleq \sqrt{\frac{2|x|}{r^2 + r x}}  \frac{d_e(N_e-1)}{\sqrt{\lambda}}\) and function $I(x)$ is defined as:
    \begin{align*}
        I(x) \triangleq \frac{1}{x}&  \Bigg |\left[C\left(x+\frac{x z_0}{(N_e-1)d_e}\right)-C\left(\frac{x z_0}{(N_e-1)d_e}\right)\right] + \\
        & \jmath \left[S\left(x+\frac{x z_0}{(N_e-1)d_e}\right)-S\left(\frac{x z_0}{(N_e-1)d_e}\right)\right]\Bigg | ,
    \end{align*} 
    where \(C(\cdot)\) and \(S(\cdot)\) are the Fresnel functions \cite[eq. (12)]{Fresnel}.
\end{Lemma}
\begin{proof}
The proof is delegated in Appendix~\ref{Prop_2_Appendix}.
\end{proof}

Function $\mathcal{K}\left(a(\pm \Delta r), \phi\right)$, defined in the previous lemma, is decreasing with respect to \(a(\pm \Delta r)\). This helps us to estimate the \(\pm\Delta r\) terms that result in a \((1-\kappa \%\)) beamforming gain loss, by simply setting \(\mathcal{K}\left(a(\pm \Delta r), \phi\right) = 0.01\kappa\) and first solving with respect to \(a(\pm \Delta r)\); the solution to this equation is denoted by\footnote{The \(a_{\kappa}\) values can be calculated offline and then used online, in contrast to \cite{SDMA_vs_LDMA} that requires numerical integration each time the azimuth angle $\phi$ changes. This happens because the function $\mathcal{K}\left(a(\pm \Delta r), \phi\right)$ has low-to-zero dependency on \(\phi\), which holds due to the fact that \(\frac{\pi^2}{90}\left(a(\pm \Delta r) \frac{(N_m-1) d_m}{2(N_e-1) d_e}|\sin(\phi)|)\right)^4 \to 0\), implying the tight approximation: \(\mathcal{K}\left(a(\pm \Delta r), \phi\right) \approx I(a(\pm \Delta r))\).} \(a_{\kappa}\). Consequently, after setting \(a(\pm \Delta r)=a_{\kappa}\) to derive \(\pm\Delta r\), the following solutions are obtained:
\begin{equation}\label{eq:Dr}
  \Delta^{\pm}_{\kappa}(r) = r^2\left(\frac{2 d_e^2 \left(N_e-1\right)^2}{\lambda a_{\kappa}^2} \mp r\right)^{-1}.
\end{equation}
The latter solutions can used to derive the depth of the BS beam focusing for which there is less than \((1-\kappa \%)\) loss, i.e., \(\frac{G_{\hat{r},\phi}}{G_{\rm opt}}\geq \kappa \%\) for \(\hat{r}\in [r-\Delta^{-}_{\kappa}(r), r + \Delta^{+}_{\kappa}(r)]\). It is noted that the term \(d_e(N_e-1)\) in \eqref{eq:Dr} is the DMA's length along the \(z\)-axis. Considering that \(N_e \! \!\gg \! \! N_m\), the DMA's aperture is approximately given as \(D \approx (N_e-1)d_e\), hence, $\Delta^{\pm}_{\kappa}(r)$ solutions can be rewritten with respect to the Rayleigh distance, \(r_{\rm RD}\), as \(\Delta^{\pm}_{\kappa}(r) \approx \frac{r^2  }{\frac{r_{\rm RD}}{ a_{\kappa}^2} \mp r }\). Note that, as \(r \!\to\! r_{{\rm lim},\kappa}\!\triangleq\! \frac{2 d_e^2 \left(N_e-1\right)^2}{\lambda a_{\kappa}^2} \!\!\approx \!\!\frac{r_{\rm RD}}{a^2_{\kappa}}\), yields \(\Delta^{+}_{\kappa}(r) \to \infty\). This resembles the limiting distance from which there ceases to exist a \(\Delta^{+}_{\kappa}(r)\), such that for \(\hat{r} > r + \Delta^{+}_{\kappa}(r)\) yields \(\frac{|\mathbf{a}^{\rm H}(r,\phi)\mathbf{a}(\hat{r},\phi) |^{2}}{N^2 } < \kappa\% \), since the UE is already in the far-field and moves further away from the BS. On the other hand, the solution \(\Delta^{-}_{\kappa}(r)\) always exists, since it indicates the direction of movement towards the BS, i.e., towards its near-field zone.
\begin{figure}
    \centering
    \includegraphics[width = \columnwidth]{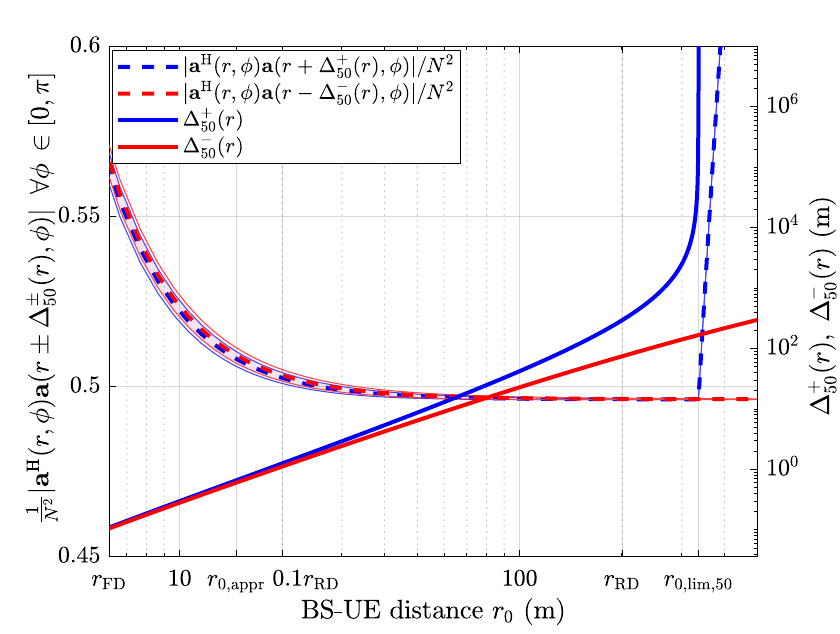}
    \caption{The relative beamforming gain for only range mismatch derived in Lemma~\ref{prop_Dr} (left vertical axis) and the range mismatch limits \(\Delta^{\pm}_{\kappa}(r)\) in \eqref{eq:Dr} for $\kappa=50$ (right vertical axis) as functions of the BS-UE distance $r_0$ in meters, considering a DMA at the BS with $N_e=200$ and $N_m = 10$ (i.e., $N=2000$ elements), $d_e=d_m=\lambda/2$ with $\lambda = 1$ cm,  and $z_0 = 1$ m. }
    \label{fig:Dr_limits}
    \vspace{-2mm}
\end{figure}

The implications of Lemma~\ref{prop_Dr} are illustrated in Fig.~\ref{fig:Dr_limits} considering $\kappa=50$ and a DMA at the BS with $N_e=200$ and $N_m = 10$ (i.e., $N=2000$), $d_e=d_m=\lambda/2$ with $\lambda = 1$ cm,  and $z_0 = 1$ m. In the left vertical axis, we have numerically evaluated the correlation factor \(\frac{|\mathbf{a}^{\rm H}(r,\phi)\mathbf{a}(r \pm \Delta^{\pm}_{50}(r),\phi) |^{2}}{N^2 }\) \(\forall\phi \in [0, \pi]\), validating that the \(\Delta^{\pm}_{50}(r)\) limits are irrespective to \(\phi\) with a maximum of \(0.5 \%\) relative error at \(r_0 = r_{\rm FD}\), which quickly converges to \(0\) as \(r_0\) increases, verifying that \(\mathcal{K}(a(\pm \Delta^{\pm}_{50}(r)),\phi) \approx I(a(\pm \Delta^{\pm}_{50}(r)))\). It is noted that, as \(r_0\) increases, the Taylor approximation in \eqref{eq:Dist_approx} holds more tightly, and hence, our analytical results become closer to the equivalent simulations, with an average relative error of \(0.6 \%\) after \(r_0=r_{0,{\rm appr}}\) and a maximum value of approximately \(10 \%\) at \(r_0 = r_{\rm FD}\). It can be also observed from the figure that, after the value \(r_{0,{\rm lim},\kappa} \triangleq \sqrt{ r^2_{{\rm lim},\kappa} + \left(z_0 + 0.5(N_e-1)d_e\right)^2 }\), \(\Delta^{+}_{\kappa}(r) \to \infty\) and the correlation factor exceeds \(0.01\kappa\). It is finally shown from the right vertical axis in Fig.~\ref{fig:Dr_limits} that \(\Delta^{+}_{50}(r)\) has a steeper rate of change compared to \(\Delta^{-}_{50}(r)\) with respect to \(r_{0}\), indicating that the depth of focus intervals become increasingly asymmetrical as \(r_{0}\) increases to the point where \(\Delta^{+}_{50}(r)\to \infty\) after \(r_{0}\geq r_{0,{\rm lim},50}\). This behavior is explained mathematically from the \(\Delta^{\pm}_{\kappa}(r)\) expression in \eqref{eq:Dr}, and its physical interpretation is that \(\Delta^{-}_{\kappa}(r)\) depicts the direction of UE movement towards the BS, where the near-field effects are stronger, hence, resulting in shorter depth of beam focus limits. 

Differently from \cite{NF_tutorial} which suggests that the $3$-dB depth of focus for linear arrays in the near field does not exist for distances \(r_0 \geq 0.1 r_{\rm RD}\), Fig.~\ref{fig:Dr_limits} showcases that, for our DMA-based planar array case, there exists this depth of focus even for \(r_0 \geq r_{\rm RD}\). To align our results with \cite{NF_tutorial}, we can set \(z_0=-0.5 (N_e-1)\) (thus, making \(r_0=r\)), yielding \(I(x)\geq 0.5\) for \(x \leq 3.12\), which results in \(\frac{r_{\rm RD}}{3.12^2}\approx 0.1 r_{\rm RD}\); a fact that further validates our analysis.  In addition, compared to \cite{SDMA_vs_LDMA}, where the authors have similarly investigated the correlation factor function with planar arrays, we provide an alternative expression tailored to our system model and derive the depth of focus. It is finally noted that the depth of focus for planar arrays, and in particular RISs, has also been investigated in \cite{Bjornson_dist}, but the provided results hold for a receiver lying along the direction dictated by the normal vector originating from the center of the metasurface.

\begin{figure}
    \centering
    \includegraphics[width=\columnwidth]{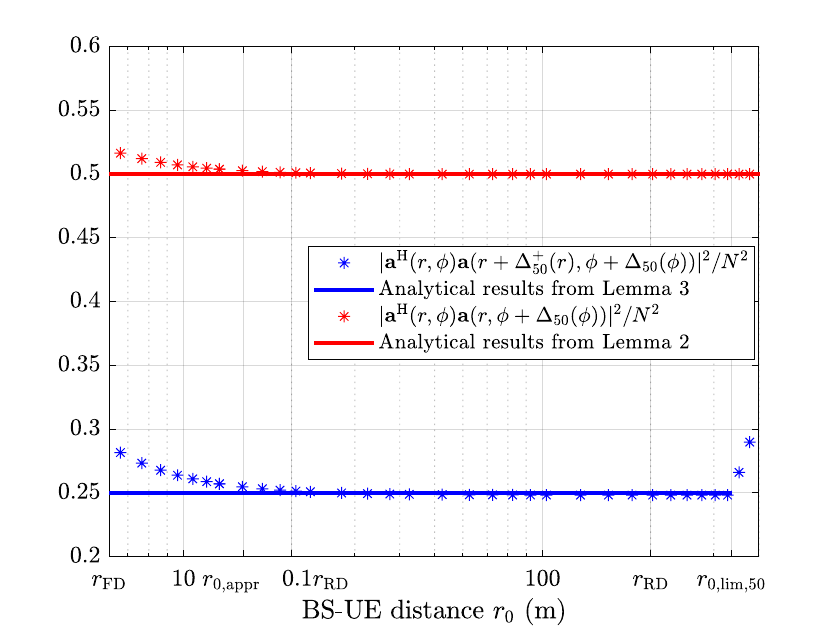}
    \caption{The relative beamforming gains for only azimuth angle mismatch (Lemma~\ref{prop_Dphi}) and for both range and angle mismatches (Lemma~\ref{prop_dr_dphi}) versus the BS-UE distance $r_0$ in meters for $\phi=\pi/4$ and the DMA parameters at the BS in Fig.~\ref{fig:Dr_limits}.}
    \label{fig:Experimental_prop_3_4}
    \vspace{-2mm}
\end{figure}

In the following, we derive similar beamforming loss bounds considering a mismatch in the azimuth angle, i.e., \(\hat{\phi}=\phi \pm \Delta \phi\).

\begin{Lemma}\label{prop_Dphi}
    When the BS beam focuses at the point \((r,\hat{\phi})\), the relative beamforming gain is derived as~\(\frac{G_{r,\hat{\phi}}}{G_{\rm opt}}\triangleq\frac{|\mathbf{a}^{\rm H}(r,\phi)\mathbf{a}(r  ,\phi \pm \Delta \phi) |^{2}}{N^2 } \approx \mathcal{L}^2(\zeta(\pm \Delta \phi))\), where function \(\mathcal{L}(x) \triangleq \frac{|\sin(x)|}{|N_m\sin(\frac{x}{N_m})|} \approx \frac{|\sin(x)|}{|x|} \) with \(\zeta(x) \triangleq N_m \frac{\pi d_m}{\lambda} \left(\cos(\phi)-\cos(\phi+x)\right)\).
\end{Lemma}

\begin{proof}
    The proof is provided in Appendix~\ref{Prop_3_Appendix}.
\end{proof}
Using the previous Lemma, we can derive the range in which the beamforming gain is at least \(\kappa \%\) of its optimum value. To this end, we solve \(\mathcal{L}^2(\zeta(\pm \Delta \phi))=0.01{\kappa} \) with respect to $\zeta(\pm \Delta \phi)$ to obtain the solution \(\zeta_{\kappa}\). Note that function \(\mathcal{L}(\cdot)\) is an even function that decreases monotonically until its first nulling point \cite[Chap. 6.3]{balanis2016antenna}, also \(\zeta(\pm \Delta \phi)\approx \pm \zeta(\Delta \phi)\) for small \(\Delta \phi\), thus, \(\mathcal{L}(\zeta(\pm \Delta \phi)) \approx \mathcal{L}(\zeta(\Delta \phi))\). If there is more than one solution, we select the first positive solution \(\zeta_{\kappa}\). Subsequently, we set \(\zeta(\pm \Delta \phi)=\zeta_{\kappa}\) and the following solution is obtained\footnote{For more accurate results, and to avoid the singularities at the points \(\phi = \{0,\pi\}\), one can numerically solve \(|\cos(\phi)-\cos(\phi \pm \Delta \phi)|=\frac{\zeta_k \lambda}{\pi N_m d_m}\) with respect to \(\Delta \phi\) without using the Taylor approximation in \eqref{eq:Dphi}.}:
\begin{equation}\label{eq:Dphi}
    \Delta_{\kappa} (\phi) = \left| \frac{\zeta_{\kappa}\lambda }{\pi  N_m d_m  \sin(\phi)}\right|.
\end{equation}
The latter solution can be now used to derive the angle width for which it holds that \( \frac{G_{r,\hat{\phi}}}{G_{\rm opt}} =\frac{|\mathbf{a}^{\rm H}(r,\phi)\mathbf{a}(r  ,\hat{\phi}) |^{2}}{N^2 } \geq \kappa \%\) for \(\hat{\phi} \in [\phi - \Delta_{\kappa} (\phi) , \phi + \Delta_{\kappa} (\phi)]\).

We next present a closed-form approximation for the relative beamforming gain for simultaneous mismatches in the range \(r\) and the azimuth angle \(\phi\).

\begin{Lemma}\label{prop_dr_dphi}
    For BS beam focusing at the point \((\hat{r},\hat{\phi})\), the relative beamforming gain is~\(\frac{G_{\hat{r},\hat{\phi}}}{G_{\rm opt}}\triangleq\frac{|\mathbf{a}^{\rm H}(r,\phi)\mathbf{a}(r \pm \Delta r ,\phi \pm \Delta \phi) |^{2}}{N^2 } \approx \mathcal{M}^2(a(\pm \Delta r),\zeta(\pm \Delta \phi)) ,\) with \(\mathcal{M}(x,y) \triangleq I(x) \mathcal{L}(y)  \).
\end{Lemma}
\begin{proof}
    See Appendix~\ref{Prop_4_Appendix}.
\end{proof}
The analytical relative beamforming gain expressions for the case studies in Lemmas~\ref{prop_Dphi} (\(\Delta_{\kappa} (\phi)\) mismatch) and~\ref{prop_dr_dphi} (\(\Delta^{\pm}_{\kappa}(r)\) and \(\Delta_{\kappa} (\phi)\) mismatches) are validated in Fig.~\ref{fig:Experimental_prop_3_4} for \(\kappa=50\) and the same DMA configuration at the BS as in Fig.~\ref{fig:Dr_limits}. It is again shown that our analytical results are sufficiently tight for \(r_0 \geq r_{0,{\rm appr}}\), but also provide satisfactory approximations before this distance value. To the best of our knowledge, this is the first work that provides a decoupled analytical function in both the range \(r\) and the azimuth angle \(\phi\) for the beam focusing gain achieved from a DMA-based (planar array) transmitter (Lemma~\ref{prop_dr_dphi}). The latter result will assist us in the following to derive the worst-case direction of the UE movement with respect to the near-field beamforming performance.

\subsection{Effective Beam Coherence Time}\label{subsec:coh_time}
We next first derive the direction of steepest descent of the relative beamforming gain, which is then utilized to evaluate the minimum distance from the BS yielding a specific relative beamforming loss. As will be shown, the latter enables the calculation of the minimum time needed for the UE to experience a specific loss relative to the theoretical gain \(G_{\rm opt}\), which will be defined as the \textit{effective beam coherence time}. 

\subsubsection{Direction of the Fastest Beamforming Gain Deterioration}
Assume a UE movement in the \(xy\)-plane of \(c\) meters from its last position \(\mathbf{p}_{t-1}=[r_{t-1},\phi_{t-1}]\) at time $t-1$. Following the cosine law, the new UE position will be \(\mathbf{p}_t=\left[r_{t-1}+d, \phi_{t-1} \pm {\rm acos}\left(1-\frac{c^2-d^2}{2r^2+2rd}\right)\right]\) with \(|d|\leq c\). Obviously, for \(d=0\), the maximum phase shift \(|\mathcal{D} \phi_{\max}|={\rm acos}\left(\frac{2r^2-c^2}{2r^2}\right)\) takes places, while for \(d=\pm c\), yields \(|\mathcal{D} r_{\max}|=c\). By expressing the values \(a(x)\) and \(\zeta(y)\) with respect to \(d\), following the respective definitions in Lemmas~\ref{prop_Dr} and~\ref{prop_Dphi}, with \(x = d\) and  \(y(d) \triangleq {\rm acos}\left(1-\frac{c^2-d^2}{2r^2+2rd}\right) \), we can re-express function \(\mathcal{M}(\cdot,\cdot)\) in Lemma~\eqref{prop_dr_dphi} for a given \(c\) as follows:
\begin{equation}\label{eq:BF_gain_for_movement_d}
\mathcal{M}(d) =
\begin{cases}
    I(a(d))\,\mathcal{L}(\zeta(y(d))), &  d \in (0,c] \\
    I(a(d))\,\mathcal{L}(\zeta(y(d))), & d \in [-c,0]
\end{cases}.
\end{equation}
This two-branch formulation stresses $\mathcal{M}(d)$'s discontinuity at \(d=0\) due to the definition of \(a(d)\) in Lemma~\ref{prop_Dr}. As a result, we can determine the direction that yields the maximum beamforming gain deterioration, by finding the critical points as: \textit{i}) the roots of the first derivative \(\frac{{\rm d}\mathcal{M}(d)}{{\rm d}d}=0\) in the second branch; and \textit{ii}) the edges of the interval \([-c,0]\). This holds due to the fact that, \(a(-|d|)>a(|d|)\Rightarrow I(a(-|d|))<I(a(|d|))\) as well as \(\zeta(y(-|d|))>\zeta(y(|d|)) \Rightarrow \mathcal{L}(\zeta(y(-|d|)))<\mathcal{L}(\zeta(y(-|d|)))\), indicating that the minimum exists in the second branch. By testing all the critical points, we can acquire \(d_{\min}\triangleq\min_{d} \mathcal{M}(d)\) subject to \(|d|\leq c\), which gives us the new position \(\mathbf{p}_{t,\min}=\left[r_{t-1}+d_{\min}, \phi_{t-1} \pm {\rm acos}\left(1-\frac{c^2-d_{\min}^2}{2r^2+2rd_{\min}}\right)\right]\), indicating the direction of the fastest deterioration, i.e., from the UE movement between positions \(\mathbf{p}_{t-1}\) and \(\mathbf{p}_{t,\min}\).

 \subsubsection{Minimum Distance Covered for \((1-\kappa\%)\) Beamforming Gain Loss}
We are now interested in finding the minimum distance needed, \(c_{\min,\kappa\%}\), so that the beamforming gain drops to \(\kappa \%\) of its optimum value \(G_{\rm opt}\). We begin by using \eqref{eq:Dr} and \eqref{eq:Dphi} to acquire \(\Delta^{-}_{\kappa}(r)\) and \(\Delta_{\kappa} (\phi)\), respectively, and then, set\footnote{Following the chord-length formula, the distance covered for a fixed range \(r\) and a phase shift \(\Delta \phi\) is given by \(2r\sin\left(0.5\Delta \phi\right)\).} \(c_{\kappa\%} \triangleq \min \{\Delta^{-}_{\kappa}(r),2r\sin\left(0.5\Delta_{\kappa} (\phi)\right)\}\). For the latter \(c_{\kappa \%}\) value, we determine the minimum value of \(\mathcal{M}(d)\). If \(\mathcal{M}(d_{\min})=\sqrt{\kappa \%}\), then \(d_{\min}=\{-c_{\kappa \%},0\}\), yielding \(c_{\min,\kappa \%}=c_{\kappa \%}\). Otherwise, we perform a bisection search for the points \(\mathbf{p}'\) along the line segment from \(\mathbf{p}_{t-1}\) to \(\mathbf{p}_{t,\min}\) (which is the point corresponding to \(d_{\min}\)), to find $\mathbf{p}'_{\kappa \%}$ solving \(  \mathcal{M}(\mathbf{p}'_{\kappa \%})=\sqrt{\kappa\%}\), and then, compute \(c_{\min,\kappa \%}=||\mathbf{p}_{t-1}-\mathbf{p}'_{\kappa \%}||_2\). For simplicity, and to avoid time/energy-consuming calculations, such as numerical solvers and one-dimensional searches, we can assume that\footnote{This typically holds true since $d_{\rm min}$ usually lies at the edges of the interval, i.e., \(d_{\rm min}=\{-c,0\}\). To demonstrate this, assume that \(\Delta^{-}_{\kappa}(r)<2r\sin(0.5\Delta_{\kappa} (\phi))\). Then, \(\mathcal{M}(d)\) decreases faster with respect to \(\Delta r\) rather than \(\Delta \phi\). For increasing \(d \in [-c,0]\), \(\mathcal{L}(\zeta(y(d)))\) decreases. However, \(I(a(-|d|))\) increases, and as assumed, \(I(\cdot)\) has a steeper rate of change, thus, leading to higher \(\mathcal{M}(d)\) values.} \(c_{\min,\kappa \%} = \min \{\Delta^{-}_{\kappa}(r),2r\sin\left(0.5\Delta_{\kappa} (\phi)\right)\}\). This enables the direct utilization of the analytical expressions \eqref{eq:Dr} and \eqref{eq:Dphi}, which yield exceptional results, as will be demonstrated in Section~\ref{sec:Results} with the performance evaluation results. 

Using the previous elaboration and the notation \(u\) for the norm of the UE velocity, we define the \textit{effective beam coherence time} as \(T_{{\rm c},\kappa \%}\triangleq \frac{c_{\min,\kappa\%}}{u}\), indicating the minimum time needed so that the beamforming gain drops to \(\kappa \%\) of \(G_{\rm opt}\). This metric will be next used to design our beam tracking protocol.

\section{Proposed Near-Field Beam Tracking}\label{sec:Proposed_BA}
In this section, we present our near-field beam tracking framework, including our dynamic non-uniform coordinate grid and our DMA-based analog and digital combining scheme. 
\vspace{-2mm}

\subsection{Design Objective}
Using the estimation \(\mathbf{\hat{p}}_{t-1}=[\hat{r}_{t-1},\hat{\phi}_{t-1}]\) of the UE position $\mathbf{p}_{t-1}$ at time $t-1$, the time interval \(T\) elapsed from this estimation, and a rough estimation of the norm of the UE's velocity, the distance \(\hat{c}\) covered by the UE can be estimated. To this end, we consider as the searching space for the UE position estimation $\mathbf{\hat{p}}_{t}=[\hat{r}_{t},\hat{\phi}_{t}]$ the surface within the UE plane of the circle centered at \(\mathbf{\hat{p}}_{t-1}\) with a radius \(\hat{c}\), ${\rm C}\left(\mathbf{\hat{p}}_{t-1},\hat{c}\right)$. Subsequently, we define our localization objective as a beamforming gain objective, i.e., finding \((\hat{r},\hat{\phi})\) so that \(G_{\hat{r},\hat{\phi}}\approx G_{\rm opt}\). The latter can be expressed via the following optimization problem:
\begin{align*} 
    \mathcal{OP}_2: \,\, \max_{\hat{r},\hat{\phi}} \,\,|\mathbf{a}^{\rm H}(\hat{r},\hat{\phi})\mathbf{h}_{\rm LoS}| \,\,\, \text{s.t.} \,\,\, \hat{r},\hat{\phi} \in {\rm C}\left(\mathbf{\hat{p}}_{t-1},\hat{c}\right).
\end{align*}
It is noted that, due to the considered DMA-based receiver at the BS, $\mathcal{OP}_2$ cannot be directly formulated from the received pilots \(\mathbf{y}_b\) in \eqref{eq:DMA_UE}. To treat this issue, we propose to receive the UE pilots through different beam focusing matrices \(\mathbf{Q}\), as will be next elaborated in Section~\ref{subsec:Hybrid_Combining}, similar to the dynamic simultaneous orthogonal matching pursuit framework in \cite[Alg.~3]{CS_DoA}, with one dominant spatial support. Furthermore, we consider that the UE coordinates maximizing the gain \(|\mathbf{a}^{\rm H}(\hat{r},\hat{\phi})\mathbf{h}|\) correspond to the LoS path, since we will beam focus in specific zones at each estimation, and scatterers residing outside those zones do not affect the search process. For scatterers residing within the area of interest ${\rm C}\left(\mathbf{\hat{p}}_{t-1},\hat{c}\right)$, we assume that their power strengths are always lower than the LoS path (due to the larger path loss), i.e., we assume that solving \(\max_{\hat{r},\hat{\phi}}  \; |\mathbf{a}^{\rm H}(\hat{r},\hat{\phi})\mathbf{h}|\) yields the UE's coordinates. 

\vspace{-2mm}
\subsection{Beam Focusing Grid and Sweeping}
To solve $\mathcal{OP}_2$ efficiently, we next focus on finding the coordinates $(\hat{r},\hat{\phi})$ resulting to a satisfactory portion of \(G_{\rm opt}\), and capitalize on the beamforming analysis in Section~\ref{sec:BF_analysis} to design a dynamic grid resulting from the time-varying continuous area of interest ${\rm C}\left(\mathbf{\hat{p}}_{t-1},\hat{c}\right)$ that includes the possible UE positions. %

\subsubsection{Dynamic Non-Uniform Coordinate Grid}\label{subsec:Dynamic_sampling}
The sampling resolution, when searching for the UE coordinates solving $\mathcal{OP}_2$, depends on how close we wish to reach the optimum beamforming gain. As described in the proposed Alg.~\ref{alg:sampling_proc}, to achieve \(\delta\%\) of \(G_{\rm opt}\), one needs to sample ${\rm C}\left(\mathbf{\hat{p}}_{t-1},\hat{c}\right)$ with this resolution, yielding approximately \(2\Delta^{+}_{\delta} (r_{i})\) distance between two consecutive radial samples \(r_i\) and \(r_{i+1}\), with \(r_{i+1}\geq r_i\) (Steps $12$ and $13$). Similarly, the angular distance between two consecutive \(\phi\) samples \(\phi_j\) and \(\phi_{j+1}\) is approximately \(2\Delta_{\delta} (\phi_{j})\) (Steps $5$ and $6$).
Specifically, each sampling point \((r_i,\phi_j)\) has a decision area such that, for all \(\hat{r}\) and \(\hat{\phi}\) within this area, it holds that \(\frac{G_{\hat{r},\phi_j}}{G_{\rm opt}}, \, \frac{G_{r_i,\hat{\phi}}}{G_{\rm opt}}\geq \delta \%\). Note that the ensemble of the sampling points has been chosen so that their decision areas cover the whole area of interest. In this way, the sampling resolution is dynamically adjusted relative to each UE position estimate $\mathbf{\hat{p}}_{t-1}$ and its estimated covered distance \(\hat{c}\). 

\begin{figure}
    \centering
    \includegraphics[width=1\columnwidth]{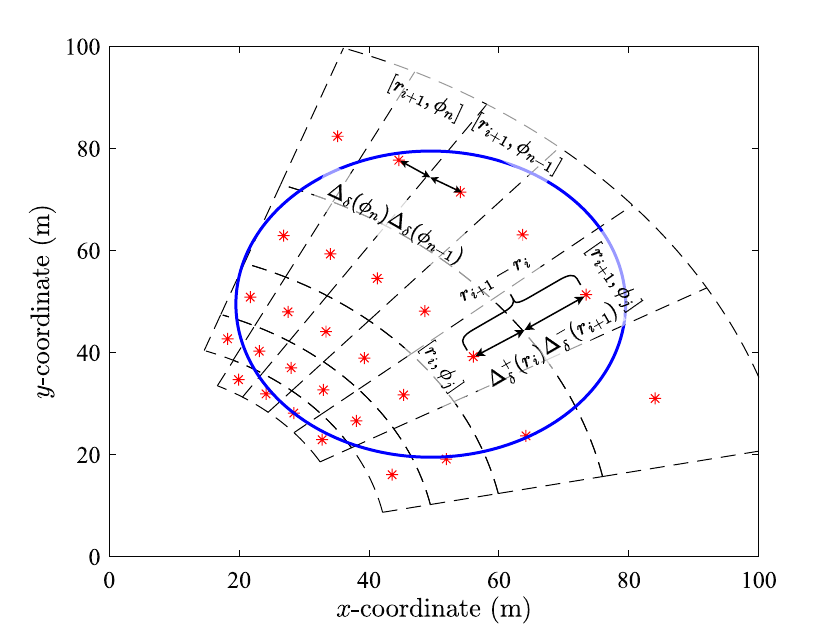}
    \caption{The proposed non-uniform coordinate grid on the $xy$ plane where the UE moves, as implemented via Alg.~\ref{alg:sampling_proc}, for a central point at the polar coordinates \(\hat{r}=70\) ${\rm m}$ and \(\hat{\phi} =\pi/4\), resolution \(\delta\%=80\%\), and radius \(\hat{c}=30\) ${\rm m}$. It is noted that, since \(\sin(\phi_j) \to 0\) when \(\phi_j \to 0\), the angular distances become wider in that regime, following the derived angular limits in \eqref{eq:Dphi}, while for the presented radial distances in \eqref{eq:Dr}, \(\Delta^{\pm}_{\delta}(r_i)\) increases with increasing $r$.}\vspace{-5mm}
    \label{fig:sampling}
\end{figure}

In contrast to the non-uniform sampling method for near-field channel estimation in~\cite{Appendix_approx}, which holds only with uniform linear arrays and lacks a dynamic dictionary or reconfigurable sampling resolution, Alg.~\ref{alg:sampling_proc} adopts the  sampling resolution to the UE area of interest ${\rm C}\left(\mathbf{\hat{p}}_{t-1},\hat{c}\right)$ and provides a non-uniform coordinate grid. These factors pronounce its application to near-field conditions. In addition, the number of sampling points and, thus complexity, are reduced in comparison with localization schemes including fixed-resolution line searches (e.g.,~\cite{DMA_loc_Nir,RIS_localisation_George_henk_ML_estimator}) and Compressive Sensing (CS) algorithms demanding fine-tuning on the coarsely estimated positions through line search maximization \cite{CS_NF_RIS}. Note that, to achieve the same results with fixed step size in sampling points, we would have to sample with the minimum sampling distance both UE coordinates.
It is also noted that, with the \(\delta \%\) resolution appearing in Alg.~\ref{alg:sampling_proc}, we define the gain percentage achieved upon estimation and not the column coherence, which in CS-based frameworks refers to the inner product of the dictionary columns \cite{CS_DoA}. In our approach, the dynamic dictionary is a matrix whose columns are the focusing vectors \(\mathbf{a}(r_i,\phi_j)\) with \((r_i,\phi_j)\) \(\forall i,j\) being the coordinates computed from Alg.~\ref{alg:sampling_proc}. Moreover, the maximum squared column coherence of our dynamic dictionary is less than \(\delta \%\) since adjacent positions in either \(\phi\) or \(r\) have \(2 \Delta_{\delta} (\phi)\) and \(2 \Delta^{+}_{\delta} (r)\) (for increasing \(r\)) differences, respectively, as shown in Fig.~\ref{fig:sampling}. The overall complexity of Alg.~\ref{alg:sampling_proc} is \(\mathcal{O}(S_r + S_\phi)\) with $S_r$ and $S_\phi$ being the total number of grid points for the radial and angular coordinates, respectively.

\begin{algorithm}[!t]
\caption{Proposed Dynamic Non-Uniform Coordinate Grid}\label{alg:sampling_proc}
 \textbf{Input:} \((\hat{r}_{t-1},\hat{\phi}_{t-1})\), \(\hat{c}\), and desired localization resolution \(\delta \%\).\\ \vspace{-3 mm}
\begin{algorithmic}[1]
\State Compute \(\mathcal{D} \phi_{\max}={\rm acos}\left(\frac{2\hat{r}^2_{t-1}-\hat{c}^2}{2\hat{r}^2_{t-1}}\right)\).
\State Initialize \( s_{\phi} = 1\) and \(\phi_{s_{\phi}} = \hat{\phi}_{t-1}-\mathcal{D} \phi_{\max}\).
\While{\(\phi_{s_\phi}+\Delta_\delta(\phi_{s_\phi}) \leq \hat{\phi}_{t-1}+ \mathcal{D} \phi_{\max}\)}
\State  \(s_\phi = s_\phi +1\).
\State  \(\phi_{\rm temp}= \phi_{s_{\phi}-1} + \Delta_\delta(\phi_{s_{\phi}-1})\).
\State  \( \phi_{s_\phi} =\phi_{\rm temp} + \Delta_\delta(\phi_{{\rm temp}})\).
\EndWhile
\State Set $S_\phi=s_\phi$ and define \(\bm{\phi}=[\phi_{1},\phi_2,\dots,\phi_{S_\phi}]\).
\State Initialize \(s_r = 1\) and \(r_{s_r} = \hat{r}_{t-1} - \hat{c}\).
\While{\(r_{s_r} + \Delta_{\delta}^{+}\!\left(r_{s_r}\right) \leq \hat{r}_{t-1} + \hat{c}\)}
\State \(s_{r} = s_r+1\).
\State \(r_{\rm temp} = r_{s_r-1} + \Delta^{+}_{\delta}\! (r_{s_r-1})\).
\State \(r_{s_r} = r_{\rm temp} + \Delta^{+}_{\delta}\!\left(r_{\rm temp}\right) \).
\State Compute \(\mathcal{D} \phi_{\max, s_r} =  {\rm acos}\left( \frac{r^2_{s_r}+ \hat{r}^2_{t-1}-\hat{c}^2}{2 r_{s_r} \hat{r}_{t-1}} \right)\).
\State Formulate \(\boldsymbol{\phi}_{s_r}\!=\!\big\{\boldsymbol{\phi}\!\mid\! [\phi_{i}\!+\!\Delta_{\delta}(\phi_i) \!\geq\!\hat{\phi}_{t-1}\!-\!\mathcal{D}\phi_{\max, s_r}] \;\!\cap\)

\hspace{-4mm}
\(\; [\phi_{i}\!-\!\Delta_{\delta}( \phi_i) \!\leq\!  \hat{\phi}_{t-1}\!+\!\mathcal{D} \phi_{\max, s_r}],\,i=1,2,\ldots,S_\phi\big \}\).
\State Compute \(S_{\boldsymbol{\phi}_{s_r}} = {\rm length}\left(\boldsymbol{\phi}_{s_r}\right)\).
\EndWhile
\State Set $S_r=s_r$.
\end{algorithmic}
\textbf{Output:} Coordinate grid \((r_{s_r},[\boldsymbol{\phi}_{s_r}]_i)\) \(\forall s_r\! =\! 1,2,\ldots, S_r\) and \(\forall i\! =\! 1,2,\ldots,S_{{\boldsymbol{\phi}}_{s_r}}\) for time $t$.
\end{algorithm}

\subsubsection{Hybrid Analog and Digital Beam Scanning}\label{subsec:Hybrid_Combining}
As depicted in Fig.~\ref{fig:sampling}, to sample the UE circular area of interest at each estimation phase, we grid the surface in arches of equal \(r\) using Alg.~\ref{alg:sampling_proc}, and then, search over the grid points for the \(\phi\) parameter within each arch. By inspecting~\eqref{eq:Dist_approx}, it can be seen that this gridding process can be implemented over-the-air by setting the analog combiner to focus on a given range \(r\), and then, digitally searching over the \(\phi\) angular samples. Specifically, the analog combiner \(\mathbf{\bar{Q}}_{r}\) and digital combiner \(\mathbf{v}_{\phi}\) that jointly focus on \((r,\phi)\) are given \(\forall i,n\) as follows:
\begin{align}
    & [\mathbf{{\bar{Q}}}_{r}]_{(i-1)N_e+n,n} \!=\! \frac{\jmath \! + \! e^{-\jmath \left(\!k \!\left(\! \frac{i_x^2d_m^2+n^2 d_e^2}{2r} + \frac{z_0 n d_e}{r} \right) {\rm -}\beta \rho_{i,n}\right)}}{2}, \label{eq:Q_r}\\
    & [\mathbf{v}_{\phi}]_i = e^{\jmath k \left(i_x d_m\cos(\phi) + \frac{i_x^2 d_m^2}{2r}\cos^2(\phi)\right)}.\label{eq:v_phi}
\end{align}
Note that, by omitting the waveguide propagation term within the receive DMA, yields \(\mathbf{v}_{\phi}^{\rm H}\mathbf{\bar{Q}}^{\rm H}_r= 0.5\mathbf{a}^{\rm H}(r,\phi)\). We employ this beam focusing process in our localization framework, i.e., change the analog sensing matrix \(\mathbf{\bar{Q}}_{r}\) to focus on different ranges \(r\), yielding the measurement \(\mathbf{y}_{b,r}\triangleq\mathbf{\bar{Q}}_{r}^{\rm H}\mathbf{h}x_u + \mathbf{\bar{Q}}_{r}^{\rm H} \mathbf{n}_b\) via \eqref{eq:DMA_UE}, and then, scan digitally for the \(\phi\) value maximizing the term\footnote{We make the typical, in the channel estimation approaches, assumption that the UE's location remains constant throughout the estimation process~\cite{Atiq2021duplex,CS_DoA,CS_NF_RIS}.} \(|\mathbf{v}^{\rm H}_\phi\mathbf{y}_{b,r}|\), thus, solving \(\mathcal{OP}_2\).

To reduce the overhead per UE position estimation when the searching radius \(\hat{c}\) for Alg.~\ref{alg:sampling_proc} is relatively large, CS can be deployed to create a dictionary from focusing vectors that are orthogonal both in angular and radial supports~\cite{Appendix_approx,CS_NF_RIS,Position_Est_mmWave}. To this end, one can first run Alg.~\ref{alg:sampling_proc} with a lower \(\bar{\delta}\% \ll 1\) resolution to compute the UE coordinates \([\bar{r},\bar{\phi}]\) yielding the maximum correlation, and then, perform a refinement with a higher resolution \(\delta'\%\) in the area defined by\footnote{This sampling process can be carried out using Alg.~\ref{alg:sampling_proc} without restricting the search to a circular space, i.e., by omitting Steps $14$ and $15$.} \(r\in [\bar{r}-\Delta^{-}_{\bar{\delta}  } (\bar{r}), \bar{r} + \Delta^{+}_{\bar{\delta}  } (\bar{r})]\) and \(\phi \in [\bar{\phi} - \Delta_{\bar{\delta}  } (\bar{\phi}), \bar{\phi} + \Delta_{\bar{\delta}  } (\bar{\phi})]\). The latter part of the procedure provides an off-grid estimation, which can also be carried out via higher complexity algorithms (e.g., via likelihood function maximization)~\cite{DMA_loc_Nir,RIS_localisation_George_henk_ML_estimator,Appendix_approx}.

\subsubsection{Initialization} 
When an estimate on the area where the UE is lying is missing, the near-field beam sweeping needs to cover the whole space \(\phi \in [0, \pi]\) and \(r_0 \in [r_{\rm FD},+\infty]\). To treat such cases, we propose the following procedure:
 \begin{itemize}
    \item \underline{Angular Space Search:} Starting from \(\phi_1=0\), we update the sampling points \(\phi_i\) via the iteration \(\phi_{i+1}={\rm acos}\left(\cos(\phi_i) -\frac{\lambda}{N_m d_m}\right)-\phi_{i}\) using the roots of \(\mathcal{L}(\cdot)\) function in~\cite[eq.~(6.11)]{balanis2016antenna}. Alternatively, for \(\phi\neq\{0,\pi\}\), we can utilize \eqref{eq:Dphi} and set \(\phi_{i+1}=\phi_i + \frac{\lambda}{N_m d_m \sin(\phi_i)}\) until \(\phi_{i +1 } > \pi\). This procedure generates orthogonal adjacent sampling points, i.e., \(|\mathbf{a}^{\rm H}(r,\phi_i)\mathbf{a}(r,\phi_{i+1})|=0\) $\forall$$i$.
     \item \underline{Radial Space Search:} Using the radial search in Alg.~\ref{alg:sampling_proc} for \(r_0 \in [r_{\rm FD},\infty]\) and low localization resolution \(\bar{\delta}\%\), yields \(|\mathbf{a}^{\rm H}(r_i,\phi)\mathbf{a}(r_{i+1},\phi)|\approx 0\) $\forall$$i$. Note that the sampling process for \(r_0 <\infty\) is equivalent to \(r_0 \leq r_{0,{\rm lim},\bar{\delta}}\), since \(\Delta^{+}_{\bar{\delta}}(r)\to \infty \) for \(r_0 = r_{0,{\rm lim},\bar{\delta}}\). This implies that no additional sampling points are needed.   
     \item  \underline{Resolution Refinement:} Search in \(r\in [\bar{r}-\Delta^{-}_{\bar{\delta}  } (\bar{r}), \bar{r} + \Delta^{+}_{\bar{\delta}  } (\bar{r})] \cup \; \phi \in [\bar{\phi} -\frac{\lambda}{N_m d_m \sin(\bar{\phi})}, \bar{\phi} + \frac{\lambda}{N_m d_m \sin(\bar{\phi})}]\). 
 \end{itemize}

\vspace{-2mm}
\subsection{Tracking Protocol and Algorithm}
\begin{figure}
    \centering
    \includegraphics[width=1\columnwidth]{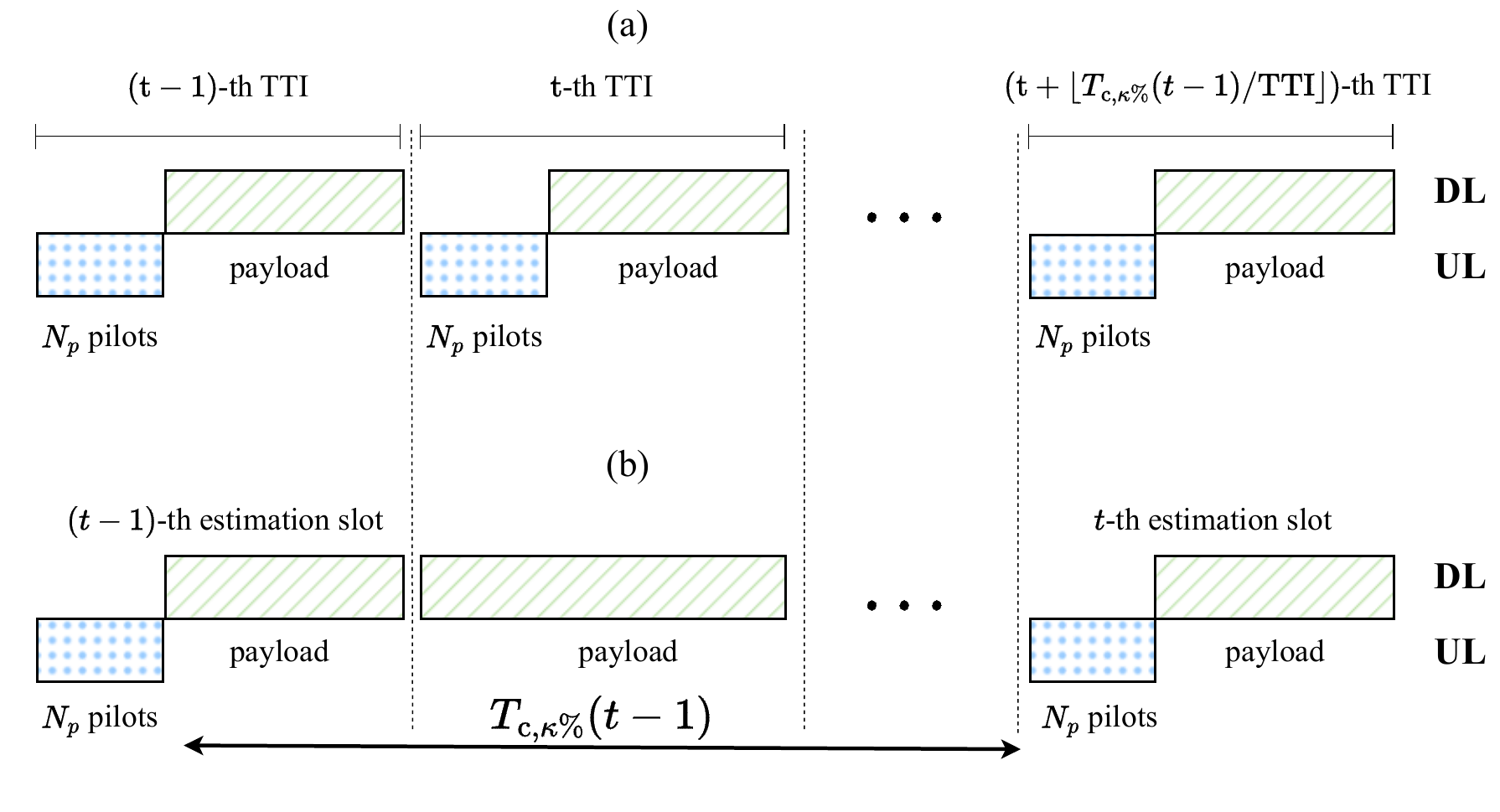}
    \caption{UL and DL in TDD: (a) Conventional communication protocol where channel estimation is performed per TTI; and (b) Proposed communication protocol where channel estimation takes places according to the effective beam coherence time, which changes dynamically depending on the beamforming loss parameter $\kappa$. In this example, \(T_{{\rm c},\kappa\%}(t-1)\) represents the time elapsing between the $(t-1)$-th and $t$-th consecutive estimation slots with the proposed near-field beam tracking protocol, the former taking place during the $({\rm t}-1)$-th TTI and the latter during the $({\rm t}+\floor{T_{{\rm c},\kappa\%}(t-1)/{\rm TTI}})$-th TTI.}
    \label{fig:UL_protocol}
    \vspace{-2mm}
\end{figure}

As discussed in Section~\ref{subsec:coh_time}, when the BS performs a UE position estimation to design its hybrid beam focusing configuration, it can also compute the minimum time needed for the DL beamforming gain to reach \(\kappa\%\) of its optimum value; this was defined as the effective beam coherence time \(T_{{\rm c}, \kappa \%}\). We propose to employ this information from the BS to request UE pilot transmissions in the UL, to proactively initiate near-field beam sweeping, as described before in Section~\ref{subsec:Hybrid_Combining}. In this way, the beam focusing sweeping takes place according to $T_{{\rm c}, \kappa \%}$, which can be designed via \(\kappa\) to depend on the link's Quality of Service (QoS), and not at every Transmission Time Interval (TTI) neither through an outage acknowledgement from the UE side, as in conventional Time Division Duplexing (TDD) communication systems. The proposed near-field beam tracking protocol is illustrated in Fig.~\ref{fig:UL_protocol}. As will become apparent in the sequel, the value of \(T_{{\rm c}, \kappa \%}\) can be different between pairs of consecutive estimation intervals. Note also that, dense pilot signaling will take place when the UE is close to the BS and the beamforming gain is significantly sensitive to position errors, while, when the UE moves further away from the BS, only a rough position estimate is needed. In this case, a small deviation from the focusing position has a negligible effect on the link's QoS. This behavior is attributed to the strong dependence of \(\mathcal{M}(\cdot)\) in~\eqref{eq:BF_gain_for_movement_d} on \(r\) for a movement of \(c\) meters. Specifically,  for \(r\gg c \), \(a(d)\to 0\) yielding \(I(a(d))\to 1\) and \(y(d) \to 0\) \(\forall d \), resulting in \(\mathcal{L}(\zeta(y(d))) \to 1\), hence, \(\mathcal{M}(d)\to 1\). 

To compute \(T_{{\rm c},\kappa \%}\), the magnitude of the UE velocity needs to be known (see Section~\ref{subsec:coh_time}). However, in typical scenarios, this quantity is neither known nor constant in time. One way to estimate this velocity magnitude $u_t$ at each estimation slot $t$ (see Fig.~\ref{fig:UL_protocol} for the connection between our method's estimation time slots and the typical TTIs) is through a geometrically weighted moving average scheme with $\gamma >1$ and $w_j \triangleq \frac{\gamma^j}{\gamma^{t}-1}$: 
\begin{equation}\label{eq:predict_velocity}
    \hat{u}_t = \max\left\{\sum_{j=0}^{t-1} w_j \hat{u}_j,u_{\rm th}\right\},
\end{equation}
where \(\hat{u}_{j+1}=||\mathbf{\hat{p}}_{j+1} - \mathbf{\hat{p}}_{j}||_2/T_{{\rm c},\kappa \%}(j) \) with $T_{{\rm c},\kappa \%}(j)$ being the effective beam coherence time computed at the end of each $j$-th estimation slot, and \(u_{\rm th}\) is used to ensure that $\hat{u}_t$ is always non-zero. Alg.~\ref{alg:tracking} summarizes our near-field beam tracking method for every $t$-th estimation slot; it is tasked to compute the BS DMA analog combiner $\mathbf{\bar{Q}}_t$ and digital beamformer $\mathbf{v}_t$, the UE position and velocity magnitude estimates \(\mathbf{\hat{p}}_t\) and \(\hat{u}_t\), respectively, as well as the effective beam coherence $T_{{\rm c},\kappa\%}(t)$ that will trigger the $(t+1)$-th estimation slot. To account for errors in the prediction and estimation of the velocity, we have used the parameters $e_c$ and $e_u$ to increase the searching radius and the velocity from \(c\) to \(c(1+e_c)\) and \(u_t\) to \(u_t(1+e_u)\), respectively. Note that, to attain the initial estimations when no prior knowledge is available, conventional estimation schemes with higher complexity can be employed (e.g.,~\cite{Wideband_Hybrid_Tracking}). In Alg.~\ref{alg:tracking}, the total number of analog combiners is equal to the number of radial samples \(S_r\). Hence, the UE area of interest lies within a radius of \(c_{\kappa \%}\), and as \(r_0\) increases, \(c_{\kappa \%}\leq \Delta^{-}_{\kappa}(r) \) increases as well. The following remark proves that, with the proposed non-uniform coordinate grid in Alg.~\ref{alg:sampling_proc}, there exists an upper bound for \(S_r\).  

\begin{algorithm}[!t]
\caption{Proposed Near-Field Beam Tracking}\label{alg:tracking}
  \textbf{Input:} Localization resolution \(\delta\%\), percentage of optimum beamforming gain \(\kappa\%\), previous estimates \(\hat{u}_{0},\ldots,\hat{u}_{t-1}\), \(\mathbf{\hat{p}}_{t-1}\), and \(T_{{\rm c},\kappa\%}(t-1)\), $e_c$ and $e_u$, and total number of pilots \(N_{\rm p}\).
\begin{algorithmic}[1]
\State The time interval \(T_{{\rm c},\kappa\%}(t-1)\) from the previous $(t-1)$-th estimation slot elapsed, request UE pilots' transmission.
\State Set the radius of the UE area of interest as $\hat{c}\!=\!c_{\kappa \%}(1+e_c)$.
\State Run \textbf{Alg. 1} using $\hat{r}_{t-1}$, $\hat{\phi}_{t-1}$, $\hat{c}$, and $\delta\%$, to obtain \(\left(r_{s_r},[\boldsymbol{\phi}_{s_r}]_i\right)\,\forall s_r=1,2,\dots,S_r,\) and \(\forall i=1,2,\dots, S_{\boldsymbol{\phi}_{s_r}}\). 
\State Perform the initializations \(\mathbf{g}=\mathbf{0}_{S_r\times1}\) and \(\mathbf{K}=\mathbf{0}_{S_r \times 2}\).
\State Set the averaging per \(\mathbf{\bar{Q}}\) configuration to \(M = \floor{N_{\rm p}/S_r}\).
\For{$s_r = 1,2,\ldots,S_r$} 
    \State Obtain $\mathbf{\bar{Q}}_{r_{s_r}}$ via \eqref{eq:Q_r} to steer on range \(r_{s_r}\). 
    \State Collect the measurements $\mathbf{y}_{b,r_{s_r}}[1],\ldots,\mathbf{y}_{b,r_{s_r}}[M]$ as 
    
    \hspace{-0.29cm} in~\eqref{eq:DMA_UE} for different noise realizations.
    \State Compute the average $\bar{\mathbf{y}}_{b,r_{s_r}}=M^{-1}\sum_{m=1}^{M}\mathbf{y}_{b,r_{s_r}}[m]$.
    \State Calculate \(i_{\max} = \arg\max_i |\mathbf{v}^{\rm H}_{[\bm{\phi}_{s_{r}}]_i}\mathbf{\bar{y}}_{b,r_{s_r}}|^2\) and set 
    
    \hspace{-0.29cm} $[\mathbf{g}]_{s_r} \!=\! |\mathbf{v}^{\rm H}_{[\bm{\phi}_{s_{r}}]_{i_{\max}}}\mathbf{\bar{y}}_{b,r_{s_r}}|^2$ and \([\mathbf{K}]_{s_r,:}\! =\! \left[r_{s_r},[\bm{\phi}_{s_{r}}]_{i_{\max}}\right]\).
    
\EndFor
\State Set \(s_{{r}_{\max}}\! = \!\arg \max_{i} [\mathbf{g}]_{i}\) and \(\mathbf{\hat{p}}_t=[\hat{r}_t,\hat{\phi}_t] \! = \! [\mathbf{K}]_{s_{r_{\max}},:}\).
\State Update the velocity estimate as \(\hat{u}_{t} = \frac{||\mathbf{\hat{p}}_{t} - \mathbf{\hat{p}}_{t-1}||_2}{T_{{\rm c},\kappa \%}(t-1)} \).
\State Compute the velocity estimate \(\hat{u}_{t+1}\) using \eqref{eq:predict_velocity} and the previous estimates \(\left(\hat{u}_{0},\hat{u}_{1},\ldots,\hat{u}_{t}\right)\).
\State \(c_{\kappa \%}\!=\!\min\!\left\{\!|2\hat{r}_t\sin\left(0.5\Delta_{\kappa}(\hat{\phi}_t) \right)\!|,{\!\Delta^{-}_{\kappa}\!(\hat{r}_t) }\!\right\}\) via \eqref{eq:Dr},~\eqref{eq:Dphi}. 
\State Compute the effective beam coherence
time for triggering the $(t+1)$-th estimation slot as \(T_{{\rm c},\kappa\%}(t) \!= \!\frac{c_{\kappa \%}}{\hat{u}_{t+1}(\!1+e_u\!)}\). 
\end{algorithmic}
\textbf{Output:} $\mathbf{\bar{Q}}_t=\mathbf{\bar{Q}}_{\hat{r}_t}$, $\mathbf{v}_t=\mathbf{v}_{\hat{\phi}_t}$, \(\mathbf{\hat{p}}_t\), \(\hat{u}_t\), and \(T_{{\rm c},\kappa\%}(t)\).
\end{algorithm}

\begin{Lemma}\label{lemma_Sr}
When sampling, using Alg.~\ref{alg:sampling_proc}, a circular area centered at the point \((r,\phi)\) with a radius \(c_{\kappa \%}\) and a localization resolution \(\delta \%\), the number of radial samples is upper bounded by the quantity \(S_{r,{\rm UB}}\leq \eta_{\kappa , \delta } + \frac{r a^2_{\kappa}}{2 r_{\rm RD}}(\eta_{\kappa , \delta } -1 ) + 1\) where \(\eta_{\kappa, \delta} \triangleq {a^2_{\kappa}}/{a^2_{\delta}}\) and \(a_{\kappa}\) as well as \(a_{\delta}\) are defined in Section~\ref{suubsec:BF_gain_mmismatch}.
\end{Lemma}
\begin{proof}
    The proof is provided in Appendix~\ref{Remark_Appendix}.
\end{proof}
Interestingly, it can be shown experimentally that the bound $S_{r,{\rm UB}}$ becomes tighter for \(r \ll \frac{r_{\rm RD}}{a^2_{\kappa}}\), since the sampling steps are nearly uniform in that regime (i.e., \(\Delta^{\pm}_{\delta} (r) \approx \Delta^{\pm}_{\delta} (r\pm c_{\kappa \%})\)). In this case, it holds \(S_{r,{\rm UB}} \approx \eta_{\kappa , \delta } + 1\). On the other hand, for \(r\geq \frac{r_{\rm RD}}{a^2_{\kappa}}\), the sampling resolution is significantly non-uniform, and consequently, $S_{r,{\rm UB}}$ becomes quite loose compared to the actual values of \(S_r\), which are again approximately given as \(S_r \approx \eta_{\kappa,\delta} + 1\). Finally, for an arbitrary radius \(\hat{c}\), one can find the closest \(c_{\kappa \%}\) to it so that \(\hat{c} = (1+e_c) c_{\kappa \%}\), and as a rule of thumb, \(S_{r,e_c,{\rm UB}} \triangleq  (1+e_c) S_{r,{\rm UB}}\) can be used.

\subsubsection{Complexity Analysis} The algorithmic Steps $6$-$12$ in Alg.~\ref{alg:tracking}, that takes place at every estimation slot $t$, result in computational complexity \(\mathcal{O}(N_m \sum_{s_r=1}^{S_r} S_{\boldsymbol{\phi}_{s_r}}) < \mathcal{O}\left(N_m S_r S_{\phi}\right)\). This value scales only with the number of microstrips $N_m$, which is significantly smaller than the total number \(N\) of the DMA elements. 
In comparison, the authors in~\cite{DMA_loc_Nir} presented a DMA-based localization scheme using a likelihood maximization approach, having the complexity $\mathcal{O}(KS_r S_\phi N N_m^2)$ with \(K\) denoting the number of iterations for the convergence of \cite[Alg.~1]{DMA_loc_Nir}.

\section{Numerical Results and Discussion}\label{sec:Results}
In this section, we present simulation results for the performance evaluation of the proposed near-field beam tracking framework. We have simulated random Bézier trajectories for the mobile UE, which are widely used for smooth path generation, known for their realistic modeling of UE trajectories~\cite{Graphics_Book}. 
Each simulation scenario includes an average over $1000$ trajectories with $100$ steps and $6$ control points each \cite[Alg.~11.4.1]{Graphics_Book}. Moreover, we have configured the time unit in each trajectory so that the average UE velocity was \(10\,\)m/s. In addition, we have considered perfect prior knowledge of two UE coordinates before running Alg.~\ref{alg:tracking}, and one scatterer was positioned uniformly within the UE circular area of interest.
In the figures that follow including performance results over the distance \(r_0\), a windowed moving average was used for averaging, and as \(r_0 \) increased, the path loss denoted by \({\rm PL}\triangleq \left(\lambda/(4 \pi r_0)\right)^2\) increased as well. Indicatively, for \(r_0 \in [5, 45]\, {\rm m}\), yields \({P_u {\rm PL}}/{\sigma^2} \in [23, 4]\) dB with \(P_u=5\)~dBm and \(\sigma^2=-94\) dBm. For the rest of the simulation parameters, unless otherwise stated, we have considered: a DMA with $N_e=200$ and $N_m = 10$, i.e., $N=2000$ metamaterials, the operating frequency \(f_c=30\) GHz, $d_e=d_m=\lambda/2$ with $\lambda = 1$ cm, and $z_0 = 1$ m. Regarding the parameters of Alg.~\ref{alg:tracking}, \(N_p=200\) pilot symbols were used, the parameter for the portion of the beamforming gain was set as \(\kappa = 50\) and the percentage of \(G_{\rm opt}\) to be achieved on estimation as \(\delta =99\). For the auxiliary variables regarding the UE velocity estimation, we have used the geometrical factor \(\gamma=2\), the threshold \(u_{\rm th}=2.5\) m/s, and the error parameters \(e_c=1.5\) and \(e_u=0.5\). 
\begin{figure}
    \centering
    \includegraphics[width=\columnwidth]{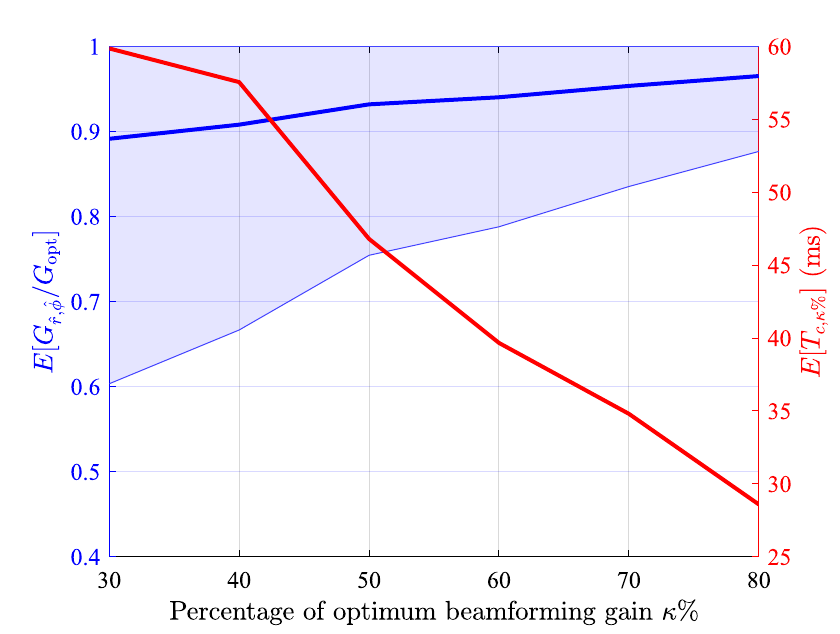}
    \vspace{-2 mm}\caption{Average relative beamforming gain \(E[G_{\hat{r},\hat{\phi}}/G_{\rm opt}]\) and its \(95\)-th percentile error bars over time with a time step of \(500\, {\rm \mu s}\) (left vertical axis in blue), as well as the average estimation time interval \(E[T_{c,\kappa \%}]\) over all estimation slots (right vertical axis in red), all as functions of the $\kappa\%$ parameter.}\vspace{-2 mm}
    \label{fig:mean_snr_over_kappa}
\end{figure}

The beamforming gain offered by the proposed near-field beam tracking scheme is investigated in Fig.~\ref{fig:mean_snr_over_kappa} for different QoS requirements indicated by the \(\kappa\%\) parameter. In particular, in the left vertical axis, the time-averaged relative beamforming gain \(E[G_{\hat{r},\hat{\phi}}/G_{\rm opt}]\) and its \(95\)-th percentile error bars (the shadowed area includes the ranges where the \(95\%\) of the collected data lies per \(\kappa\) value) are included, whereas the right vertical axis lists the \(T_{{\rm c}, \kappa \%}\) averaged over all estimation slots. It can be observed that the achievable relative beamforming gain is always greater than the \(\kappa \%\) threshold 
with \(95\%\) confidence. Furthermore, it reaches on average at least $90\%$ even for small \(\kappa \%\) values. In addition, as expected, \(E[T_{c,\kappa \%}]\) decreases with increasing \(\kappa \%\), since denser estimations need to take place over time to satisfy a stricter QoS constraint.

\begin{figure}
    \centering
    \includegraphics[width=\columnwidth]{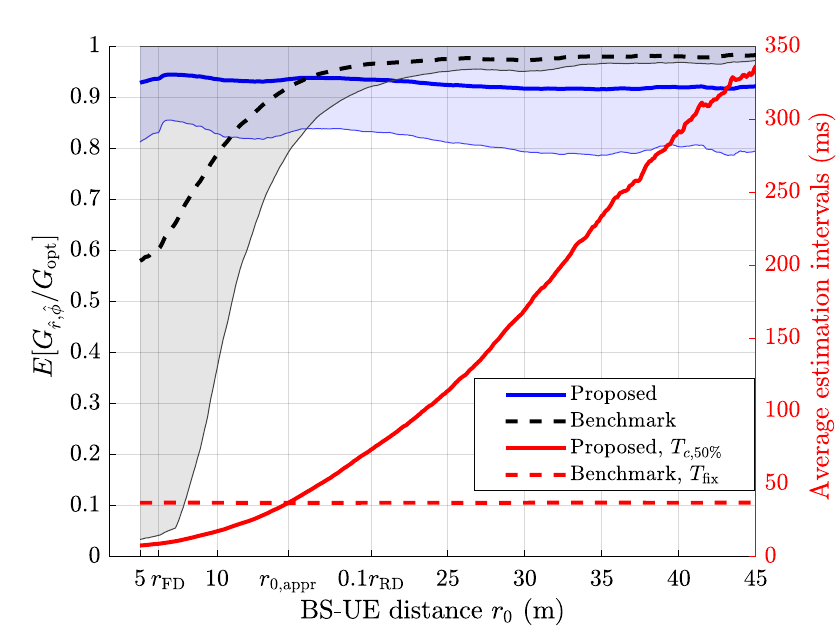}
    \vspace{-2 mm}\caption{Average relative beamforming gain \(E[G_{\hat{r},\hat{\phi}}/G_{\rm opt}]\) and its \(90\)-th percentile error bars over time with a time step of \(500\, {\rm \mu s}\) (left vertical axis). Solid blue and dashed black lines correspond respectively to the proposed tracking algorithm and a benchmark tracking scheme with the fixed parameters \(T_{\rm fix}\triangleq E[T_{{\rm c},50 \%}]\), \(\mathcal{D} r_{\rm fix}\triangleq E[\Delta^{\pm}_{99}(r)]\), and \(\mathcal{D} \phi_{\rm fix}\triangleq E[\Delta_{99} (\phi)]\). The right vertical axis in red includes the values of the average estimation intervals for both schemes. All metrics are plotted versus the BS-UE distance \(r_0\) for $\kappa=50$.}\vspace{-2 mm}
    \label{fig:fixed_comparison_avg_gain}
\end{figure}

\begin{figure}
    \centering
    \includegraphics[width=\columnwidth]{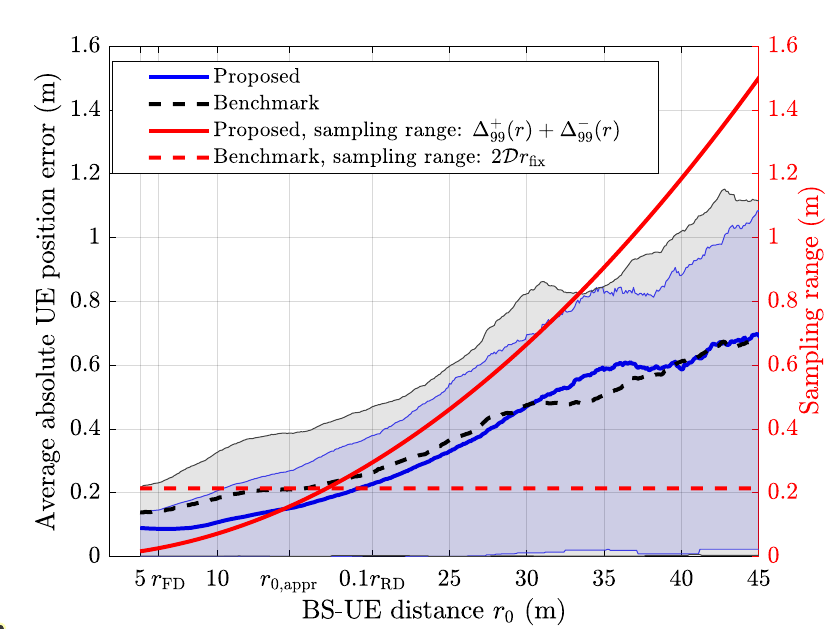}
    \vspace{-2 mm}\caption{Absolute UE position error averaged over estimation slots and its \(90\)-th percentile error bars for both tracking schemes in Fig.~\ref{fig:fixed_comparison_avg_gain} (left vertical axis). In the right vertical axis in red, the solid line indicates the sampling range of \(r\) with the proposed scheme, given as \(\Delta^{-}_{99}(r) +\Delta^{+}_{99} (r) \), whereas the dashed line corresponds to the fixed sampling range \(2\mathcal{D} r_{\rm fix}\).
    }\vspace{-2 mm}
    \label{fig:fixed_comparison_position_error}
\end{figure}

In Figs.~\ref{fig:fixed_comparison_avg_gain} and~\ref{fig:fixed_comparison_position_error}, we compare the performance of the proposed tracking scheme with a benchmark tracking scheme adopting a fixed estimation interval of duration \(T_{\rm fix} \triangleq E[T_{{\rm c},50 \%}]\) as well as the fixed sampling resolutions \(\mathcal{D}r_{\rm fix}\triangleq E[\Delta^{\pm}_{99}(r)]\) and \( \mathcal{D} \phi_{\rm fix} \triangleq E[\Delta_{99} (\phi)]\); the averages are taken over all estimation slots of the proposed scheme. In the left vertical axis of Fig.~\ref{fig:fixed_comparison_avg_gain}, \(E[G_{\hat{r},\hat{\phi}}/G_{\rm opt}]\) and its \(90\)-th percentile error bars are depicted with respect to the BS-UE distance \(r_0\), while the average estimation intervals are illustrated in the figure's right vertical axis. As shown and as expected, the estimation intervals increase with increasing \(r_0\). Interestingly, for both schemes, the average achieved beamforming gain exceeds the \(50 \%\) QoS requirement for all simulated range values, with the proposed scheme yielding always a gain larger than the \(90\%\) of its optimum value. For the benchmark scheme, this gain is stabilized after \(r_0 \geq 0.1 r_{\rm RD}\), but is highly variable for smaller distances. This fact signifies the importance of the reconfigurable estimation intervals adopted within our proposed tracking framework. As shown from the solid red curve, our scheme performs dense estimations at small \(r_0\) values and sparser ones when \(r_0\) increases. Indicatively, at \(r_0=40\)~m, \(7\) times more estimation slots are used by the benchmark scheme, as compared to the proposed one, while \(E[G_{\hat{r},\hat{\phi}}/G_{\rm opt}]\) equals respectively to \(98\%\) and \(92\%\); i.e., a slight difference of \(6\%\). 

Fig.~\ref{fig:fixed_comparison_position_error} showcases the average distance between position sampling points for increasing \(r_0\) values in conjunction with the UE position error achieved. In particular, in the left vertical axis, the absolute position error averaged over estimation slots and its \(90\)-th percentile error bars for both tracking schemes in Fig.~\ref{fig:fixed_comparison_avg_gain} are depicted, while, in the right vertical, we plot the sampling range defined as the length of the decision range for a sampling point at \(r_0\). This length is given by \(\Delta^{+}_{99}(r) +\Delta^{-}_{99}(r) \) for the proposed scheme (the decision areas are shown in Fig.~\ref{fig:sampling}) and by \(2 \mathcal{D} r_{\rm fix}\) for the benchmark. It can be seen from the figure that our scheme, capitalizing on the dynamic non-uniform coordinate grid in Alg.~\ref{alg:sampling_proc}, achieves position error (blue solid line) less than the sampling range (red solid line) for \(r_0 \geq r_{0,{\rm appr}}\). In addition, it outperforms the benchmark scheme  when sampling with smaller sampling range, i.e., for \(r_{0}< 17\) m; at this point the red solid and dashed lines meet. On the other hand, when the benchmark scheme samples with a smaller sampling range (i.e., for \(r_{0} \geq 17\, {\rm m}\)), there is no performance gain, which can be explained as follows:
\begin{itemize}
    \item By selecting the sampling resolution to achieve \(0.99G_{\rm opt}\), denser sampling yields negligible differences on the gain \(|\mathbf{v}^{\rm H}_{[\bm{\phi}_{s_{r}}]_{i_{\max}}} \mathbf{\bar{y}}_{b,r_{s_r}}|^2\) (see Steps $10$ and $12$ in Alg.~\ref{alg:tracking}). Considering the noise threshold, this implies that it is hard to distinguish between range and angular coordinates with differences less than \(\Delta^{\pm}_{99}(r) \) and \( \Delta_{99}( \phi)\), respectively. 
     \item The average position error via the benchmark scheme for $\mathcal{D} r_{\rm fix} =0.1$ m and \(\mathcal{D} \phi_{\rm fix}= 2.14^{\circ}\) (black dashed line) is larger than the fixed sampling range (red dashed line) for \(r_{0} \geq r_{0,{\rm appr}}\). This indicates that more samples are unnecessary, in the sense that they cannot be distinguished. Indicatively, at \(r_0=40\) m, the benchmark scheme samples with \(6\) times smaller range than the proposed one, but the same average position error is achieved.
    \item Erroneous previous estimations, due to undersampling for low \(r_0\) values, affect the performance of future estimations as well. This is attributed to error propagation. 
\end{itemize}

\begin{figure}
    \centering
    \includegraphics[width=\columnwidth]{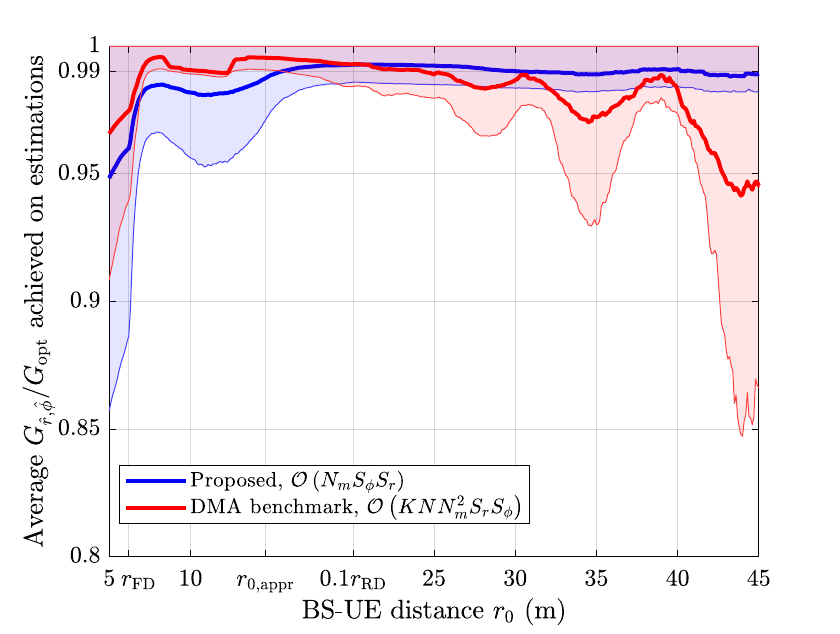}
   \vspace{-2 mm} \caption{Achieved \(G_{\hat{r},\hat{\phi}}/G_{\rm opt}\) averaged over estimation slots versus the BS-UE distance \(r_0\). For the DMA benchmark, we have used Alg.~\ref{alg:tracking} replacing Steps $4$-$12$ with the localization method in~\cite[Alg.~1]{DMA_loc_Nir}. For the latter method, the initialization point at the \(t\)-th estimation slot was set as \(\mathbf{\hat{p}}_{t-1}\) and a maximum of \(K=20\) algorithmic iterations was used.}\vspace{-6 mm}
    \label{fig:comparison_w_Nirs}
\end{figure}
\vspace{-1 mm}

In Fig.~\ref{fig:comparison_w_Nirs}, we compare the localization performance of the proposed scheme with the maximum likelihood approach presented in~\cite{DMA_loc_Nir}. We particularly plot the portion of \(G_{\rm opt}\) averaged over estimation slots versus the distance \(r_0\). Recall that this metric constitutes our localization objective in \(\mathcal{OP}_2\). %
As demonstrated in \cite[Fig.~4]{DMA_loc_Nir}, the performance of this localization scheme is sensitive to initialization points. Targeting an upper bound performance for this scheme, we have deleted divergent points resulting from bad initializations, and carried out the simulations without the presence of a scatterer; \cite{DMA_loc_Nir} assumed a pure LoS scenario. The spikes in Fig.~\ref{fig:comparison_w_Nirs} for the DMA benchmark scheme can be explained from the fact that the analog combiner \(\mathbf{Q}\) is directive both in \(r\) and \(\phi\), thus, rendering the received signal too ``biased'' and prone to bad initializations. However, with our scheme, \(\mathbf{Q}\) focuses only to a given \(r\) yielding a wider beam. It can be also observed from the figure that our scheme outperforms the higher complexity DMA benchmark for \(r_0 \geq r_{0,{\rm appr}}\). In this regime, the approximation in \eqref{eq:Dist_approx} holds tight, and our scheme achieves  \(99 \%\) of \(G_{\rm opt}\) irrespective of the \(r_0\) value. Interestingly, satisfactory results for our scheme are also demonstrated for \(r_0 < r_{\rm FD}\); in this region, larger than \(94 \%\) of \(G_{\rm opt}\) is achieved. Finally, by observing Fig.~\ref{fig:comparison_w_Nirs} in conjunction with  Fig.~\ref{fig:fixed_comparison_position_error}, it can be concluded that, while the position error increases with increasing \(r_0\), the average relative beamforming gain remains stable. This validates the goal of our near-field beam tracking framework.

\vspace{-2mm}

\section{Conclusions}
In this paper, we studied the near-field beam tracking problem in a high frequency point-to-point wireless communication system between a DMA-based BS and a mobile single-antenna UE. A theoretical analysis of the optimum achievable beamforming gain and this gain's loss due to UE coordinate mismatch was presented, overcoming limitations of previous relevant works, such as the dependency between the polar coordinate parameters in the angular and depth width derivations. Novel analytical results extending the near-field effects to larger distances, even larger than the Rayleigh distance, were also derived. In addition, we designed a dynamic non-uniform coordinate grid for effectively sampling the UE circular area of interest at each position estimation slot. Moreover, we introduced the metric of the effective beam coherence time, indicating the minimum time needed for the UE to experience a specific QoS degradation level, that was dynamically  computed at the BS during each UE position estimation slot. The performance of the proposed near-field beam tracking framework, that triggers beam focusing sweeping only when the BS estimates that its provided beamforming gain drops below a threshold from its theoretically optimum value, was extensively investigated via computer simulations. The presented theoretical analysis was validated and it was showcased that the proposed near-field beam tracking approach outperforms benchmark schemes.   
\vspace{-1mm}

 \appendices
 \renewcommand{\theequation}{A-\arabic{equation}}
  \setcounter{equation}{0}  
  
  \section{Proof of Lemma 1}
  The relative beamforming gain for a mismatch in the radial coordinate, i.e., when \(\hat{r}=r\pm \Delta r\) with $\Delta r\geq0$, is derived as:\vspace{-1mm}
  \label{Prop_2_Appendix}
  \begin{equation}
  \begin{split}
      & \frac{|\mathbf{a}^{\rm H}(r, \phi)\mathbf{a}(r \pm \Delta r, \phi)|}{N} =\\
      & \bigg|\frac{1}{N_m}\sum_{i_x=-0.5(N_m-1)}^{0.5(N_m-1)}e^{\pm\jmath k \left( \frac{i_x^2 d_m^2 \sin^2(\phi)}{2}\frac{\Delta r}{ r^2 \pm r \Delta r} 
 \right)} \\
 &\times \frac{1}{N_e}\sum_{n=0}^{N_e-1}  e^{\pm\jmath k \left(\left(\frac{n^2 d_e^2}{2} + z_0 n d_e\right)\frac{\Delta r}{ r^2 \pm r \Delta r}\right)}\bigg|.
  \end{split}
  \end{equation}
For the first sum, we make use of the approximation in \cite[Appendix~A]{Appendix_approx}, while, for the second sum, we employ the Riemann sum approach similar to \cite{SDMA_vs_LDMA}, first aligning it to our system model. By setting \(n'\triangleq \frac{n}{N_e-1}\) and \(\tau^{\pm} \triangleq \frac{\Delta r}{ r^2 \pm r \Delta r}\) yields:
\vspace{-1mm}
\begin{equation}
\begin{split}
      & \frac{|\mathbf{a}^{\rm H}(r, \phi)\mathbf{a}(r \pm \Delta r, \phi)|}{N} \approx\\
      & \bigg|\frac{2}{N_m}\int_{0}^{0.5 (N_m-1)} e^{\pm\jmath k \left( \frac{i_x^2 d_m^2 \sin^2(\phi)}{2}\tau^{\pm}  \right)}di_x  \\
      & \times \int_{0}^{1} e^{\pm\jmath k \left(\left(\frac{n'^2(N_e-1)^2 d_e^2}{2} + z_0 n'(N_e-1) d_e\right)\tau^{\pm}\right)} dn'\bigg|.
\end{split}
\end{equation}
After some algebraic manipulations and using the definitions of the Fresnel functions \(C(\cdot)\) and \(S(\cdot)\) \cite[eq. (12)]{Fresnel}, as well as the functions $a(x)$, 
\(\overline{a}(x)\triangleq a(x) \frac{(N_m-1) d_m}{2(N_e-1)d_e}|\sin(\phi)|\), and \(b \triangleq \frac{ z_0}{(N_e-1)d_e}\), we derive the following analytical approximation:\vspace{-1mm}
\begin{align}\label{eq:App_corr_fun}
     & \frac{|\mathbf{a}^{\rm H}(r, \phi)\mathbf{a}(r \pm \Delta r, \phi)|}{N}  \approx \nonumber\\
     & \left|\mathcal{D}(\overline{a}(\pm \Delta r))\right| \left|D\left(a(\pm \Delta r)(1 +b)\right)- D\left(a(\pm \Delta r)b\right)\right|,
\end{align}
where we have used the function definition:
\begin{align}
 \mathcal{D}(x) \triangleq \begin{cases}
                 \frac{C(x) + \jmath S(x)}{x}, & x\neq 0,\\
                 1,& x=0
                \end{cases},
\end{align}
the fact that \(|\mathcal{D}^{\ast}(x)|=|\mathcal{D}(x)|\), and the following first-order Taylor approximation of $|\mathcal{D}(x)|$ at $x_0\to 0$:\vspace{-1mm}
\begin{align}\label{eq:App_taylor_approx_fresnel}
 \left|\mathcal{D}(x)\right| \approx  \left(1-\frac{\pi^2}{90}x^4\right).
\end{align}\vspace{-1mm}
\vspace{-1mm}
\renewcommand{\theequation}{B-\arabic{equation}}
  \setcounter{equation}{0}  

\section{Proof of Lemma 2}
The relative beamforming gain for a mismatch in the azimuth coordinate, i.e., when \(\hat{\phi}=\phi \pm \Delta \phi\), is obtained as:\vspace{-1mm}
  \label{Prop_3_Appendix}
\begin{equation}\label{eq:App_B_corr_fun}
\begin{split}
      & \frac{|\mathbf{a}^{\rm H}(r, \phi)\mathbf{a}(r , \phi \pm \Delta \phi )|}{N} =\\
      &\bigg|\frac{1}{N_m}\sum_{i_x=-0.5(N_m-1)}^{0.5(N_m-1)}e^{-\jmath k (i_x d_m \left(\cos(\phi) - \cos(\phi \pm \Delta \phi)\right)}  \\
      & \times e^{\jmath k \left(\frac{i_x^2 d_m^2 \left(\sin^2(\phi) - \sin^2(\phi \pm \Delta \phi)\right)}{2r}\right)} \bigg|.
\end{split}
\end{equation}
It holds that \(\xi_{i_x}\triangleq\frac{i_x^2 d_m^2 \left(\sin^2(\phi) - \sin^2(\phi \pm \Delta \phi)\right)}{2r} \to 0\) \(\forall i_x\) in \eqref{eq:App_B_corr_fun}, yielding \(\frac{|\mathbf{a}^{\rm H}(r, \phi)\mathbf{a}(r , \phi \pm \Delta \phi )|}{N} = \mathcal{L}(\zeta(\pm \Delta \phi))\) where $\mathcal{L}(x)$ is defined in \cite[eq.~(6-10d)]{balanis2016antenna}.
We continue by proving that the term including \(\xi_{i_x}\) leads to negligible changes to the latter beamforming gain expression. Similar to Appendix~\ref{Prop_2_Appendix}, we approximate the sum in \eqref{eq:App_B_corr_fun} with an integral as follows:\vspace{-1mm}
  \begin{align}
      & \bigg|\frac{1}{N_m}\sum_{i_x=-0.5(N_m-1)}^{0.5(N_m-1)} \! \! \! \! e^{\jmath k \xi_{i_x}} \bigg| \! \approx \! \left|\mathcal{D}\left(w\right)\right|,
  \end{align}
where \(w \triangleq \sqrt{\frac{|\sin^2(\phi)-\sin^2(\phi \pm \Delta \phi)|}{2r\lambda}}(N_m-1)d_m\). For less than \(1 \%\) error, we set \(|\mathcal{D}(w)|^2 \geq 0.99\), which yields \(w \leq 0.46\). Then, we can equivalently write the inequality:\vspace{-1mm}
\begin{equation}\label{eq:Dphi_ineq}
    (N_m-1)^2d_m^2 \leq \frac{0.46^2 2 r \lambda }{ |\sin^2(\phi)-\sin^2(\phi \pm \Delta \phi)|}.
\end{equation}
Moreover, we make the following assumptions: \textit{i}) \( r \geq r_{\rm FD}\); and \textit{ii}) the phase differences \(\Delta \phi\) are given by \eqref{eq:Dphi}. This implies that the values of \(x\) for which \(\mathcal{L}^2(x)\) lies within its first nulling point are \(|x|\leq \pi\). Hence, after performing the approximation \(|\sin^2(\phi)-\sin^2(\phi \pm \Delta \phi)|\approx |\sin(2 \phi) \Delta \phi|\) at \(\Delta \phi \to 0\), we can rewrite \eqref{eq:Dphi_ineq} as follows:\vspace{-1mm}
\begin{equation}\label{eq:Dphi_ineq_final}
    \frac{(N_m-1)^2d_m^2}{N_m d_m D \sqrt{D}} \leq \frac{ 0.131}{\sqrt{\lambda}|\cos(\phi)|}.
\end{equation}
It can be seen that \eqref{eq:Dphi_ineq_final} holds true since \(\frac{(N_m-1)^2d_m^2}{N_m d_m D \sqrt{D}} \ll 1\), and due to the fact that, for wireless communications above $30$ GHz, the wavelength is \(\lambda \leq 1 \, {\rm cm}\), resulting in \(\frac{0.131}{\sqrt{\lambda}|\cos(\phi)|}\geq 1.31\).

\renewcommand{\theequation}{C-\arabic{equation}}
\setcounter{equation}{0}  
\section{Proof of Lemma 3}
\label{Prop_4_Appendix}
The relative beamforming gain for both a radial (\(\hat{r}= r \pm \Delta r \) with \(\Delta r\geq 0\)) and an angular (\(\hat{\phi}=\phi \pm \Delta \phi\)) mismatch, is derived as follows. Utilizing Lemmas~\ref{prop_Dr} and~\ref{prop_Dphi} and omitting the term \({i_x^2 d_m^2} \left(\frac{\sin^2(\phi)}{2r} - \frac{\sin^2(\phi \pm \Delta \phi)}{2(r\pm \Delta r)}\right)\), yields the expression:\vspace{-1mm}
\begin{align}
    &\frac{|\mathbf{a}^{\rm H}(r, \phi)\mathbf{a}(r \pm \Delta r , \phi \pm \Delta \phi )|}{N} =\nonumber\\
    & \bigg|\frac{1}{N_mN_e}\!\sum_{i_x=-0.5(N_m-1)}^{0.5(N_m-1)}e^{-\jmath k \left(i_x d_m \left(\cos(\phi) - \cos(\phi \pm \Delta \phi)\right) \right)}\\
    & \times\!\sum_{n=0}^{N_e-1}\!\!e^{\pm\jmath k \left(\!\!\left(\frac{n^2 d_e^2}{2} + z_0 n d_e\right)\frac{\Delta r}{ r^2 \pm r \Delta r}\right)}\bigg| 
    \!=\!I(a(\pm \Delta r))\mathcal{L}(\zeta(\pm \Delta \phi)).\nonumber
\end{align}
Similar to Appendix~\ref{Prop_3_Appendix}, the resulting error from omitting the previously mentioned term is negligible in the regime where:\vspace{-1mm}
\begin{equation}\label{eq:Dr_Dphi_ineq}
    (N_m-1)^2d_m^2 \leq \frac{0.46^2 2(r \pm \Delta r) \lambda }{ |\sin^2(\phi)\left(1\pm\frac{\Delta r}{r}\right)-\sin^2(\phi \pm \Delta \phi)|}.
\end{equation}
Assuming further that \(r - \Delta r \geq r_{\rm FD}\) and \(\frac{\Delta r \sin^2(\phi)}{r} \to 0\), the latter expression reduces to \eqref{eq:Dphi_ineq_final}, concluding the proof.

\renewcommand{\theequation}{D-\arabic{equation}}
\setcounter{equation}{0}
\section{Proof of Lemma 4}
  \label{Remark_Appendix}
Following Alg.~\ref{alg:sampling_proc}, two adjacent sampling points in the radial coordinate, namely $(r_i,\phi_j)$ and $(r_{i+1},\phi_j)$ $\forall$$i,j$ with \(r_{i+1}>r_i\) and \(r_{i+1},r_{i} \in [r-c_{\kappa \%}, r+ c_{\kappa \%}]\), are distanced by the quantity \(\Delta^{+}_{\delta} (r_i) + \Delta^{-}_{\delta}(r_{i+1}) \). Since \(\Delta^{+}_{\delta} (r)\) increases with respect to \(r\), the minimum distance between any two elements becomes less than \( 2 \Delta^{+}_{\delta}(r-c_{\kappa \%}) \). Consequently, by dividing the searching space into the two regions \([r-c_{\kappa \%},r]\) and \([r,r+ c_{\kappa \%}]\), the number of sampling points is upper bounded as: \(S_{r,{\rm UB}} \triangleq \frac{c_{\kappa \%}}{2 \Delta^{+}_{\delta} (r-c_{\kappa \%})} + \frac{c_{\kappa \%}}{2 \Delta^{-}_{\delta}(r)} + 1 
    \leq  \frac{\Delta^{-}_{\kappa}( r)}{ 2\Delta^{+}_{\delta} (r-c_{\kappa \%})} + \frac{\Delta^{-}_{\kappa} (r)}{ 2\Delta^{+}_{\delta} (r)} + 1\),
where we have also accounted for the fixed sampling point at \(r=r-c_{\kappa \%}\). 
After some straightforward mathematical manipulations, utilizing \eqref{eq:Dr} and the approximation \( \frac{2 d_e^2 \left(N_e-1\right)^2}{\lambda a_{\kappa}^2} \approx \frac{r_{\rm RD}}{a^2_{\kappa}} \), yields \(S_{r,{\rm UB}}\leq \eta_{\kappa , \delta } + \frac{r a^2_{\kappa}}{2 r_{\rm RD}}(\eta_{\kappa , \delta } -1 ) + 1\).

\bibliographystyle{IEEEtran}
\bibliography{references}

\begin{thebibliography}{10}
\providecommand{\url}[1]{#1}
\csname url@samestyle\endcsname
\providecommand{\newblock}{\relax}
\providecommand{\bibinfo}[2]{#2}
\providecommand{\BIBentrySTDinterwordspacing}{\spaceskip=0pt\relax}
\providecommand{\BIBentryALTinterwordstretchfactor}{4}
\providecommand{\BIBentryALTinterwordspacing}{\spaceskip=\fontdimen2\font plus
\BIBentryALTinterwordstretchfactor\fontdimen3\font minus
  \fontdimen4\font\relax}
\providecommand{\BIBforeignlanguage}[2]{{%
\expandafter\ifx\csname l@#1\endcsname\relax
\typeout{** WARNING: IEEEtran.bst: No hyphenation pattern has been}%
\typeout{** loaded for the language `#1'. Using the pattern for}%
\typeout{** the default language instead.}%
\else
\language=\csname l@#1\endcsname
\fi
#2}}
\providecommand{\BIBdecl}{\relax}
\BIBdecl

\bibitem{THz_loc_tutorial}
H.~Chen \emph{et~al.}, ``A tutorial on terahertz-band localization for {6G}
  communication systems,'' \emph{IEEE Commun. Surveys \& Tuts.}, vol.~23,
  no.~3, pp. 1780--1815, 2022.

\bibitem{Holographic_MIMO_et_al}
C.~Huang \emph{et~al.}, ``Holographic {MIMO} surfaces for \(6\){G} wireless
  networks: Opportunities, challenges, and trends,'' \emph{IEEE Wireless
  Commun.}, vol.~27, no.~5, pp. 118--125, 2020.

\bibitem{XLMIMO_tutorial}
Z.~Wang \emph{et~al.}, ``A tutorial on extremely large-scale {MIMO} for {6G}:
  {F}undamentals, signal processing, and applications,'' \emph{IEEE Commun.
  Surveys \& Tuts.}, early access, 2024.

\bibitem{NF_tutorial}
Y.~Liu \emph{et~al.}, ``Near-field communications: {A} tutorial review,''
  \emph{IEEE Open J. Commun. Society}, vol.~4, pp. 1999--2049, 2023.

\bibitem{Hybrid_mimo_tracking}
J.~Zhao \emph{et~al.}, ``Angle domain hybrid precoding and channel tracking for
  millimeter wave massive {MIMO} systems,'' \emph{IEEE Trans. Wireless
  Commun.}, vol.~16, no.~10, pp. 6868--6880, 2017.

\bibitem{SDMA_vs_LDMA}
Z.~Wu and L.~Dai, ``Multiple access for near-field communications: {SDMA} or
  {LDMA}?'' \emph{IEEE J. Sel. Areas Commun.}, vol.~41, no.~6, pp. 1918--1935,
  2023.

\bibitem{time_difference_arrival_localisation}
Y.~Wang and K.~C. Ho, ``Unified near-field and far-field localization for {AOA}
  and hybrid {AOA}-{TDOA} positionings,'' \emph{IEEE Trans. Wireless Commun.},
  vol.~17, no.~2, pp. 1242--1254, 2018.

\bibitem{RIS_aided_time_of_arrival}
K.~Keykhosravi \emph{et~al.}, ``{RIS}-enabled {SISO} localization under user
  mobility and spatial-wideband effects,'' \emph{IEEE J. Sel. Top. Signal
  Process.}, vol.~16, no.~5, pp. 1125--1140, 2022.

\bibitem{Localization_TOA_Primer}
N.~Garcia \emph{et~al.}, ``Direct localization for massive {MIMO},'' \emph{IEEE
  Trans. Signal Process.}, vol.~65, no.~10, pp. 2475--2487, 2017.

\bibitem{DMA_Magazine}
N.~Shlezinger \emph{et~al.}, ``Dynamic metasurface antennas for \(6\){G}
  extreme massive {MIMO} communications,'' \emph{IEEE Wireless Commun.},
  vol.~28, no.~2, pp. 106--113, 2021.

\bibitem{RIS_overview}
E.~Basar \emph{et~al.}, ``Reconfigurable intelligent surfaces for {6G}:
  {E}merging applications and open challenges,'' \emph{IEEE Veh. Technol.
  Mag.}, to appear, 2024.

\bibitem{stacked_hmimo}
J.~An \emph{et~al.}, ``Stacked intelligent metasurfaces enabling efficient
  holographic mimo communications for {6G},'' \emph{IEEE J. Sel. Areas
  Commun.}, vol.~41, no.~8, pp. 2380--2396, 2023.

\bibitem{DMA_1bit_uplink}
P.~Gavriilidis \emph{et~al.}, ``Metasurface-based receivers with 1-bit {ADC}s
  for multi-user uplink communications,'' in \emph{Proc. IEEE ICASSP}, Seoul,
  South Korea, 2024.

\bibitem{gavras2023duplex}
I.~Gavras \emph{et~al.}, ``Full duplex holographic {MIMO} for near-field
  integrated sensing and communications,'' in \emph{Proc. EUSIPCO}, Helsinki,
  Finland, 2023.

\bibitem{DMA_UL_mMIMO}
N.~Shlezinger \emph{et~al.}, ``Dynamic metasurface antennas for uplink massive
  {MIMO} systems,'' \emph{IEEE Trans. Commun.}, vol.~67, no.~10, pp.
  6829--6843, 2019.

\bibitem{DMA_near_field_channel}
H.~Zhang \emph{et~al.}, ``Beam focusing for multi-user {MIMO} communications
  with dynamic metasurface antennas,'' in \emph{Proc. IEEE ICASSP}, Toronto,
  Canada, 2021.

\bibitem{DMA_loc_Nir}
Q.~Yang \emph{et~al.}, ``Near-field localization with dynamic metasurface
  antennas,'' in \emph{Proc. IEEE ICASSP}, Rhodes, Greece, 2023.

\bibitem{DMA_energy_eff}
L.~You \emph{et~al.}, ``Energy efficiency maximization of massive {MIMO}
  communications with dynamic metasurface antennas,'' \emph{IEEE Trans.
  Wireless Commun.}, vol.~22, no.~1, pp. 393--407, 2023.

\bibitem{HMIMO_survey_et_al}
T.~Gong \emph{et~al.}, ``Holographic {MIMO} communications: {T}heoretical
  foundations, enabling technologies, and future directions,'' \emph{IEEE
  Commun. Surveys \& Tuts.}, vol.~26, no.~1, p. 196–257, 2024.

\bibitem{Near_Field_localization_MUSIC}
B.~Friedlander, ``Localization of signals in the near-field of an antenna
  array,'' \emph{IEEE Trans. Signal Process.}, vol.~67, no.~15, pp. 3885--3893,
  2019.

\bibitem{CRB_source_loc}
J.~Chen \emph{et~al.}, ``Maximum-likelihood source localization and unknown
  sensor location estimation for wideband signals in the near-field,''
  \emph{IEEE Trans. Signal Process.}, vol.~50, no.~8, pp. 1843--1854, 2002.

\bibitem{NF_tracking}
A.~Guerra \emph{et~al.}, ``Near-field tracking with large antenna arrays:
  Fundamental limits and practical algorithms,'' \emph{IEEE Trans. Signal
  Process.}, vol.~69, pp. 5723--5738, 2021.

\bibitem{RIS_and_NF_tracking}
S.~Palmucci \emph{et~al.}, ``Two-timescale joint precoding design and {RIS}
  optimization for user tracking in near-field {MIMO} systems,'' \emph{IEEE
  Trans. Signal Process.}, vol.~71, pp. 3067--3082, 2023.

\bibitem{RIS_localisation_George_henk_ML_estimator}
Z.~Abu-Shaban \emph{et~al.}, ``Near-field localization with a reconfigurable
  intelligent surface acting as lens,'' in \emph{Proc. IEEE ICC}, Montreal,
  Canada, 2021.

\bibitem{Ramezani2024}
P.~Ramezani and E.~Bj{\"o}rnson, \emph{Near-Field Beamforming and Multiplexing
  Using Extremely Large Aperture Arrays}.\hskip 1em plus 0.5em minus
  0.4em\relax Cham: Springer International Publishing, 2024, pp. 317--349.

\bibitem{Bjornson_dist}
E.~Bj{\"o}rnson \emph{et~al.}, ``A primer on near-field beamforming for arrays
  and reconfigurable intelligent surfaces,'' in \emph{Proc. IEEE Asilomar Conf.
  Signals, Sys., and Comp.}, Pacific Grove, USA, 2021.

\bibitem{Appendix_approx}
M.~Cui and L.~Dai, ``Channel estimation for extremely large-scale {MIMO}:
  {F}ar-field or near-field?'' \emph{IEEE Trans. Commun.}, vol.~70, no.~4, pp.
  2663--2677, 2022.

\bibitem{Near_field_NOMA}
Z.~Ding, ``Resolution of near-field beamforming and its impact on {NOMA},''
  \emph{IEEE Wireless Commun. Lett.}, to appear, 2024.

\bibitem{apostol1991calculus}
T.~M. Apostol, \emph{Calculus, Volume 1}.\hskip 1em plus 0.5em minus
  0.4em\relax John Wiley \& Sons, 1991.

\bibitem{Rayleigh_Fresnel_distances}
K.~T. Selvan and R.~Janaswamy, ``Fraunhofer and fresnel distances: Unified
  derivation for aperture antennas,'' \emph{IEEE Antennas Propag. Mag.},
  vol.~59, no.~4, pp. 12--15, 2017.

\bibitem{vlachos2019wideband}
E.~Vlachos \emph{et~al.}, ``Wideband {MIMO} channel estimation for hybrid
  beamforming millimeter wave systems via random spatial sampling,'' \emph{IEEE
  J. Sel. Topics Signal Process.}, vol.~13, no.~5, pp. 1136--1150, 2019.

\bibitem{Wideband_Hybrid_Tracking}
K.~Dovelos \emph{et~al.}, ``Channel estimation and hybrid combining for
  wideband terahertz massive {MIMO} systems,'' \emph{IEEE J. Sel. Areas
  Commun.}, vol.~39, no.~6, pp. 1604--1620, 2021.

\bibitem{RIS_energy_eff_et_al}
C.~Huang \emph{et~al.}, ``Reconfigurable intelligent surfaces for energy
  efficiency in wireless communication,'' \emph{IEEE Trans. Wireless Commun.},
  vol.~18, no.~8, pp. 4157--4170, 2019.

\bibitem{Fresnel}
J.~Sherman, ``Properties of focused apertures in the {F}resnel region,''
  \emph{IRE Trans. Antennas Propag.}, vol.~10, no.~4, pp. 399--408, 1962.

\bibitem{balanis2016antenna}
C.~A. Balanis, \emph{Antenna theory: analysis and design}.\hskip 1em plus 0.5em
  minus 0.4em\relax John wiley \& sons, 2016.

\bibitem{CS_DoA}
S.~Park and R.~W. Heath, ``Spatial channel covariance estimation for the hybrid
  {MIMO} architecture: A compressive sensing-based approach,'' \emph{IEEE
  Trans. Wireless Commun.}, vol.~17, no.~12, pp. 8047--8062, 2018.

\bibitem{CS_NF_RIS}
O.~Rinchi \emph{et~al.}, ``Compressive near-field localization for multipath
  {RIS}-aided environments,'' \emph{IEEE Commun Lett.}, vol.~26, no.~6, pp.
  1268--1272, 2022.

\bibitem{Atiq2021duplex}
M.~A. Islam \emph{et~al.}, ``Direction-assisted beam management in full duplex
  millimeter wave massive {MIMO} systems,'' in \emph{Proc. IEEE GLOBECOM},
  Madrid, Spain, 2021.

\bibitem{Position_Est_mmWave}
A.~Shahmansoori \emph{et~al.}, ``Position and orientation estimation through
  millimeter-wave {MIMO} in \(5\){G} systems,'' \emph{IEEE Trans. Wireless
  Commun.}, vol.~17, no.~3, pp. 1822--1835, 2018.

\bibitem{Graphics_Book}
M.~K. Agoston, \emph{Computer Graphics and Geometric Modelling:
  {I}mplementation \& Algorithms}.\hskip 1em plus 0.5em minus 0.4em\relax
  Springer London, 2005.

\end{thebibliography}

\end{document}